\documentclass[prb,twocolumn,amssymb,showpacs,superscriptaddress]{revtex4}
\usepackage{graphicx,amsmath,graphics}
\usepackage{color}
\usepackage{epstopdf}
\begin{document}
\title{Geometrical frustration effects on charge-driven quantum phase transitions}
\author{L. Cano-Cort\'es}
\affiliation{Departamento de F\'isica Te\'orica de la Materia Condensada,
Universidad Aut\'onoma de Madrid, Madrid 28049, Spain}
\author{A. Ralko}
\affiliation{Institut N\'eel-CNRS and Universit\'e Joseph Fourier,
Bo\^ite Postale 166, F-38042 Grenoble Cedex 9, France}
\author{C. F\'evrier}
\affiliation{Institut N\'eel-CNRS and Universit\'e Joseph Fourier,
Bo\^ite Postale 166, F-38042 Grenoble Cedex 9, France}
\author{J. Merino}
\affiliation{Departamento de F\'isica Te\'orica de la Materia Condensada,
Universidad Aut\'onoma de Madrid, Madrid 28049, Spain}
\author{S. Fratini}
\affiliation{Institut N\'eel-CNRS and Universit\'e Joseph Fourier,
Bo\^ite Postale 166, F-38042 Grenoble Cedex 9, France}
\date{\today}
\begin{abstract}
The interplay of Coulomb repulsion and geometrical frustration on charge-driven quantum phase transitions is explored.  The ground state 
phase diagram of an extended Hubbard model on an anisotropic triangular lattice relevant to quarter-filled layered organic materials 
contains homogeneous metal,  'pinball'  and three-fold charge ordered metallic phases. The stability of the 'pinball' phase occurring for  
strong Coulomb repulsions is found to be strongly influenced by geometrical frustration. 
A comparison with a spinless model reproduces the transition from the homogeneous metallic phase to a pinball liquid,  which indicates that the spin correlations should play a much smaller role than the charge correlations in the metallic phase close to the charge ordering transition.
Spin degeneracy 
is, however, essential to describe the dependence of the system on
geometrical frustration.  Based on finite temperature
Lanczos diagonalization we find that the effective Fermi temperature scale, $T^*$, of the homogeneous metal vanishes at the
quantum phase transition to the ordered metallic phase driven by the Coulomb repulsion. Above this temperature scale 'bad' metallic behavior is found which is robust 
against geometrical frustration in general.  Quantum critical phenomena are not found whenever nesting of the 
Fermi surface is strong,  possibly  indicating a first order transition instead. 'Reentrant' behavior 
in the phase diagram is encountered whenever the $2k_F$-CDW instability competes with the Coulomb driven three-fold
charge order transition. The relevance of  our results to the family of quarter-filled materials: $\theta$-(BEDT-TTF)$_2$X is discussed. 
\end{abstract}
\maketitle

\section{Introduction}

\label{sec:intro}

Strongly correlated electron materials are often characterized by
 complex phase diagrams, 
reflecting an intricate interplay between magnetic, 
orbital, lattice and charge degrees of
freedom. As these excitations couple to the conduction electrons,
the metallic  state expected in the 
absence of interactions 
has to compete 
with several ordered phases.
Examples of these  materials include cuprate superconductors, 
nickelates, heavy fermion compounds,
transition metal dichalchogenides, 
organic charge transfer salts 
and the iron based pnicitide superconductors, all presenting various
forms of magnetic, orbital and  charge order. 
Even when a metallic
phase is stabilized, these systems are generally found to exhibit
large effective mass enhancements and electrical 
resistivities violating the  Ioffe-Regel-Mott (IRM)
condition \cite{Emery,Merino00,Allen02,Hussey,Calandra,Gunnarsson03}. 
Surprisingly enough, such ``bad'' metallic behavior does not
impede the emergence of superconductivity, but rather  appears to be a
prerequisite for the achievement of high critical temperatures  
 \cite{BasovManifesto11}.

Charge ordered (CO) phases are commonly
observed in the class of 
two-dimensional organic compounds
$\theta$-ET$_2$X  (ET$=$BEDT-TTF,
bisethylenedithio-tetrafulvalene) 
\cite{Miyagawa,Yamamoto,Watanabe99,Watanabe04,Watanabe05},  and ascribed to the prominent role of electron-electron interactions \cite{Mori00}. 
At the non-interacting level,  
these compounds are predicted to be metals with $3/4$-filled electronic bands.
The observation of electronic ordering implies 
that the magnitude of electron-electron interactions is comparable with the 
widths of the relevant electronic bands constructed from the $\pi$
molecular orbitals. In turn, the presence of such strong 
interactions raises questions about the nature of the 
metallic phase in these materials, that should exhibit
distinctive features of ``correlated electron systems'' in the Mott
sense.  The proximity to charge ordering instabilities, with the
possible emergence of quantum critical points as the transition
temperature is made to vanish, is also expected to strongly alter the
physical properties of the metal. All these ingredients 
should lead to measurable deviations from
the usual Fermi liquid behavior \cite{Merino06,CanoCortes10}, 
in close analogy with heavy fermion systems 
\cite{Coleman,Gegenwart,Lohneysen,Miyake}.

The minimal theoretical description of the electronic properties of
$\theta$-ET$_2$X  organic conductors  
is based on the two-dimensional 
extended Hubbard model (EHM) on a triangular lattice.
Several theoretical studies have aimed at reproducing the different
CO patterns realized in this class of materials, either within the
framework of the EHM itself or its generalizations, including longer ranged
electronic interactions and various types of  
electron-lattice interactions \cite{Kuroki09,Udagawa07}.
In the present work we focus on the following open issues: 
(i) how does the strength of the local
  Coulomb correlations modify the nature of the metallic phase as well as 
its CO instabilities, (ii)
what are the effects of geometrical
frustration in the electron motion arising from the 
 triangular molecular arrangement, and
(iii)  how does the proximity to a given CO phase extend its influence
  onto the properties of the correlated metal, possibly
  leading to non-Fermi liquid behavior?

In Sec.~\ref{sec:model} we set the minimal electronic 
model needed for the study
of electronic properties of 
$\theta$-(ET)$_2$X compounds and provide a brief review of established
theoretical results. In Sec.~\ref{sec:T0} the model is solved by
Lanczos diagonalization and the zero-temperature phase diagram 
is obtained for different degrees of geometrical frustration. 
In Sec.~\ref{sec:finiteT} the resulting metallic phases
are explored through a finite-$T$ Lanczos  
diagonalization calculation \cite{Jaklic,Liebsch08,Liebsch09}. 
We theoretically explore  
the consequences  of a quantum critical point (QCP) 
at a charge ordering transition driven
by the quantum 
fluctuations associated with strong inter-site Coulomb repulsion.
Our results are compared with a spinless
calculation in order to assess
the importance of the magnetic degrees of freedom  
in the  observed quantum criticality and make contact with the
existing literature.
The relevance of the present results to the physics of 
$\theta$-type ET compounds
is discussed in the conclusive Sec.~\ref{sec:conclusions}.

\section{Model and method}
\label{sec:model}

\subsubsection{Extended Hubbard model on the triangular lattice}

\begin{figure}
 \center
 \includegraphics[clip,scale=0.25]{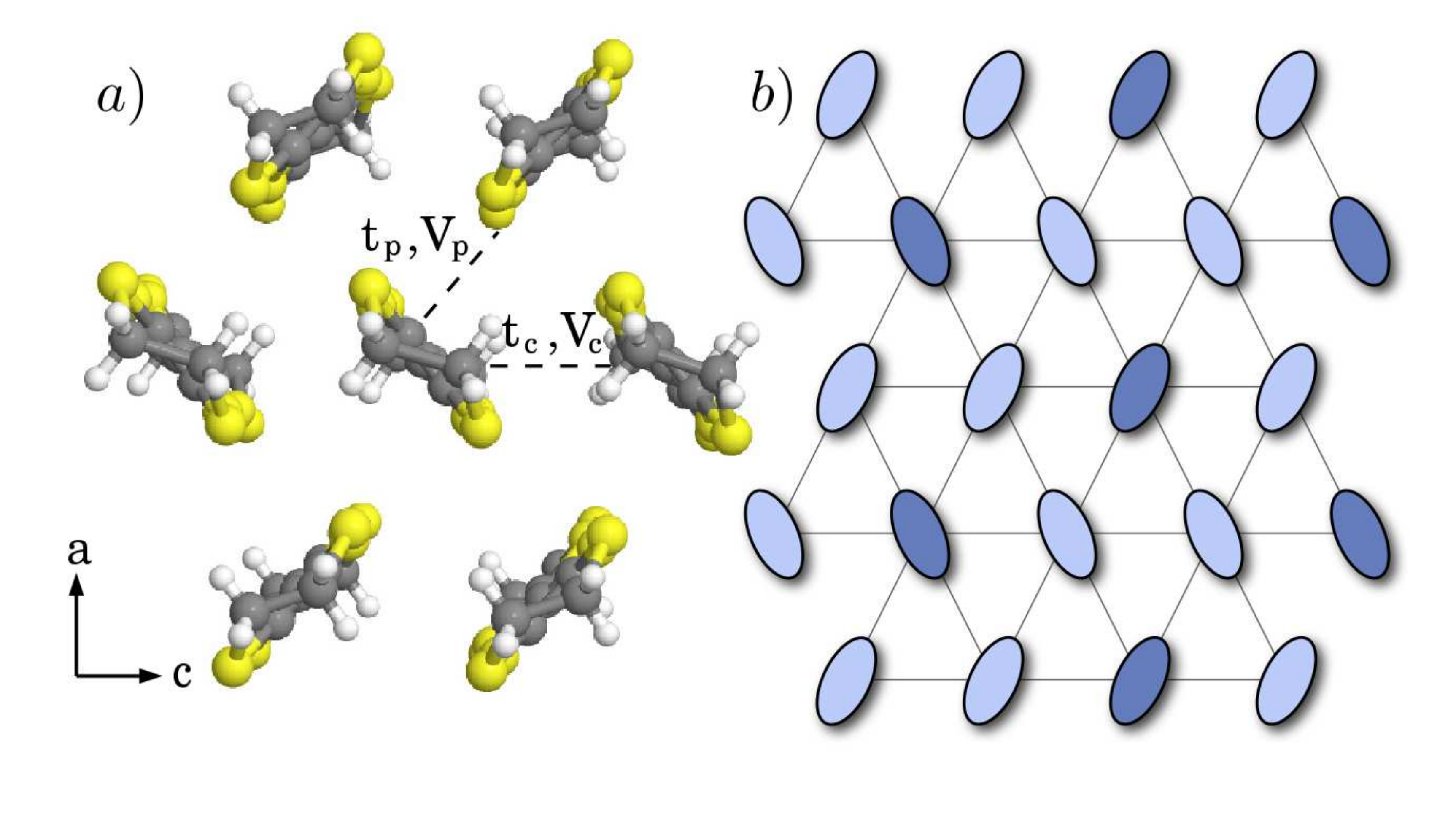}
 \caption{\label{triangular}(Color online) (a)
  Arrangement of BEDT-TTF molecules in the conducting layers of 
  $\theta$-type ET
  crystals, with the corresponding 
 transfer integrals and nearest-neighbor Coulomb interactions. (b)
 Threefold charge ordered phase in the triangular lattice considered
 in this work.    The triangular lattice is defined by the real space
   unit vectors ${\bf a}_1=(1,0)$ defining the $c$ direction and
   ${\bf a}_2=(1/2,\sqrt{3}/2)$.  }
\end{figure}

Quasi two-dimensional organic conductors (general formula $A_2B$) 
are  charge transfer compounds composed of alternating layers of 
conducting (donor) molecules $A$ and insulating (acceptor) units $B$. 
They exhibit a large variety of molecular arrangements 
corresponding to different polytypes classified by greek
characters \cite{Mori00,Seo04}.  The materials
of the  $\theta$-ET$_2$X class have a triangular lattice structure,
shown in Fig.~\ref{triangular}, with an average of 
$n=3/2$ electrons per
molecule,  fixed by complete 
charge transfer between $A$ and $B$ units. Since this
corresponds to a three-quarter filled electronic band, these materials
should be normal metals in the absence of interactions.   

The electronic properties of $\theta$-ET$_2$X materials are commonly described
 via the extended Hubbard Model (EHM),
\begin{eqnarray}
  \label{eq:EHM}
  H &=& -t_p \sum_{\langle ij \rangle _p\sigma} \left( c_{i\sigma}^{\dagger} c_{j\sigma}+H.c. \right)
  -t_c \sum_{\langle ij \rangle _c\sigma} \left( c_{i\sigma}^{\dagger} c_{j\sigma}+H.c. \right)
  \nonumber \\
  &&+ U \sum_i n_{i\uparrow}n_{i\downarrow} + V_p \sum_{\langle ij \rangle _p} n_i n_j + V_c \sum_{\langle ij \rangle _c} n_i n_j.
\end{eqnarray}
This model includes transfer integrals between the $\pi$-orbitals of
nearest-neighboring molecules in the conducting plane, labeled
by $t_p$ and $t_c$ according to the bond directions, 
a local (on-site) Coulomb repulsion energy $U$ as well as non-local
(nearest-neighbor) repulsion terms $V_c$,$V_p$ (see 
Fig.~\ref{triangular}) 
\footnote{We work explicitly with electrons.  To make contact with other
theoretical works in the literature, the case 
$t_p<0$ for $n=3/2$ is equivalent to taking $t_p>0$
for $n=1/2$ (one-quarter filling) without changing the absolute sign
of $t_c$. }.

Early mean-field calculations \cite{Kuroki06,Seo00,Kaneko} 
indicated that three types of striped patterns (vertical, diagonal and
horizontal) are realized 
 depending on the relative magnitude of the non-local Coulomb
interaction parameters. These  results
were later confirmed by more advanced numerical techniques that can properly
account for 
electronic correlations,  such as 
exact diagonalization (ED)  \cite{Clay,Merino05} and density matrix renormalization group (DMRG) 
\cite{NishimotoDMRG08}.

More interesting from our perspective
is the  isotropically interacting  case, $V_p = V_c \equiv V$
\cite{Mori03,Kaneko,Kuroki06,Kuroki09,HottaFurukawa06,Miyazaki,WatanabeVMC06,NishimotoDMRG08}.
There,  because of the  frustration of inter-molecular interactions 
induced by the triangular lattice geometry,     
an alternative  charge ordering pattern with three-fold periodicity
is favored with respect to the (degenerate) striped
arrangements, illustrated in  Fig.~\ref{triangular}b. A more exotic situation is found in the limit of 
strong  local Coulomb interactions (or, similarly, in a fully spin polarized
electron system, i.e. for 
spinless electrons), where the constraint of no double occupancy
on molecular sites converts this threefold order
into  a partially ordered phase termed  "pinball liquid''
(PL)
\cite{HottaFurukawa06,WatanabeVMC06,NishimotoDMRG08,CanoCortes10}: 
this state shows a three-sublattice structure with the same symmetry
as the threefold phase, 
in which the carriers of one sublattice are essentially localized
as a Wigner-crystal (pin), with the remaining charges (balls) forming 
an itinerant liquid on the interstitials.  
It is not clear at present how 
the transition between these two qualitatively
different forms of threefold order takes place as a function of  
the local Coulomb repulsion $U$. This issue
will therefore be thoroughly discussed here.

In addition to the effects of the local electronic correlations, we
are interested in the effects of geometrical frustration in the
electronic motion, that are triggered by the 
strongly directional $\pi$-overlaps between neighboring molecules 
\footnote{In this work we consider the maximally
  frustrated  case concerning interactions ($V_p=V_c=V$), while
  varying the degree of frustration in the electron motion via the
  parameter $t_c/t_p$.  
  Previous studies of the EHM on the square lattice correspond, in the
  present language, to the case 
  $V_p=V$ and $V_c=0$.}.
This issue is of particular importance to actual
materials, as the relative values of the transfer integrals $t_c$ and
$t_p$ can be tuned experimentally by applying pressure or by
 chemical substitution, which modifies the relative angles between
 neighboring molecules \cite{Mori,McKenzie2001}. 
As a general observation, negative values of the ratio $t_c/t_p$ 
produce the highest charge ordering temperatures 
\cite{Mori,Kuroki09}, while vanishing or positive values lead to
glassy  (X$=$CsCo(SCN)$_4$, X$=$CsZn(SCN)$_4$) \cite{Monceau} or even 
superconducting (X$=$I$_3$) ground states \cite{Mori,Kuroki09,Tamura94}.
From a more theoretical point of view, how the
system evolves from a perfectly isotropic triangular
lattice at $t_c=\pm t_p$ to a square
lattice at $t_c=0$ remains an open issue.

Transfer integrals obtained through the H\"uckel approximation 
in quarter-filled  $\theta$-ET crystals 
are in the range:  $-0.5 \lesssim t_c/t_p \lesssim 1.5$, with 
$t_p \approx -0.05-0.1$ eV 
\cite{McKenzie2001}. These values generally differ from the ones extracted
from optical reflectivity  
and de Haas Van Alphen experiments \cite{Tamura94} for each specific
crystal. On the other hand,  Coulomb repulsion energies in organic
molecular 
crystals \cite{Canocortes07} have been estimated by calculating the
screening corrections to the bare repulsion energies of the  
isolated molecules, $U_0$ and $V_0$,  obtained from {\it ab initio}
calculations \cite{Scriven09}.  These calculations lead
to Hubbard parameters of the model Eq. (\ref{eq:EHM}):  $U \sim U_0/2 
\sim (15-20)|t_p|$, and $V_p \sim V_c \sim U/2$, with a 
 bandwidth $W \sim (8-9)|t_p|$. These Coulomb energies are larger than assumed in previous works \cite{Merino05,Seo06,WatanabeVMC06,NishimotoDMRG08}:  $U \sim (8-10)|t_p|$, and
  $V \sim (1-3)|t_p|$, as extracted from optical reflectivity measurements \cite{Mori00}. 
 The degree of uncertainity in the microscopic 
parameters implies that a general understanding of the 
 model  and its phase diagram in the full parameter space $U$, $V$ and $t_c/t_p$ is  essential. This is 
 the main focus of the present work.  

\subsubsection{Finite-$T$ Lanczos approach}

We perform ED calculations through a finite-$T$ Lanczos algorithm with periodic
boundary conditions \cite{Jaklic,Liebsch08}.   The large number of excited states inherent to the
many-body problem which are needed to evaluate statistical sums is cutoff by keeping only a small 
number of low lying states at each temperature.  
This is performed through an Arnoldi algorithm \cite{Liebsch09} which reduces the size of the
Hilbert space enormously. The accuracy of the method is restricted to temperatures which are 
not too low, i.e. not lower than the energy of the lowest excitation of the quantum many-body system. For the method
to be practical $T$ should not be too large that one needs to keep too many states in
the statistical sums.   Finite-size effects are somewhat reduced by the effect of temperature and
the method is quite reliable for  extracting integrated
properties. Instead, spectral properties such
as optical conductivity and photoemission spectra are 
prone to large
finite size effects, and will not be analyzed here.  

Due to the high computational demand of the finite-$T$
algorithm, the calculations are performed on
an $N_s=12$ site cluster.  In principle, a larger $N_s=18$ cluster  whose  geometry is
also suitable for reproducing the three-fold CO pattern could be 
used  at $T=0$. However, we shall not consider such case because 
a $3/4$-filling implies a different number
of spin up and spin down electrons, while the
ground state is expected to be in a $S=S_z=0$ state.

We  characterize the 
physical properties of the different phases based on the following
quantities, that are accessible through finite-$T$ Lanczos  calculations. 

(i) {\em Charge correlation function}: 
The charge structure factor signalling the possible occurrence of a charge ordered state in 
the system is evaluated at finite-$T$ through: 
\begin{eqnarray}
  \label{eq:Cq}
  C({\bf q})=  \frac{1}{Z}
  \sum_{m} e^{-\beta E_m} \langle m | \frac{1}{N_s^2}  \sum_{i,j} e^{i{\bf q}\cdot {\bf R}_{ij}}n_i n_j | m \rangle.
\end{eqnarray}
Here $Z=\sum_m e^{-\beta E_m} $ is the partition function of the system and
$\beta=1/k_BT$. 
A charge ordered state with modulation ${\bf Q}$  is signalled if $C({\bf Q})$ is finite
 in the thermodynamic limit. 
The three-fold ordering corresponds to a charge density 
modulation with wavevector  ${\bf Q}=(2\pi/3,2\pi/\sqrt{3})$, which 
lies at the corner of the hexagonal Brillouin zone, see Fig. \ref{fig:FS} 
(all corners are equivalent, being connected by reciprocal lattice vectors
or time-reversal symmetry).   
An accurate numerical determination of the phase boundaries
should rely on 
a proper finite-size scaling of the results. While this 
is prohibitive for the fermionic system under study 
due to the rapidly increasing size of the Hilbert space, 
the ordering transitions can still be identified 
as the locus of steepest variation of charge 
correlations upon varying the microscopic parameters of the model.

(ii) {\em  Kinetic energy}: 
This quantity provides direct information on how the motion of the
charge carriers is slowed down by interactions. 
It can be evaluated with high accuracy from the finite-$T$ Lanczos
diagonalization, because it results from a quantum mechanical and thermal average over
a huge number of states.  By normalizing it to  a reference 
non-interacting value $K_0$, it gives valuable information on the degree of
electronic correlations in the many-body  system \cite{Millis04,Qazilbash09}.
Under suitable assumptions, this quantity can be
compared with optical absorption experiments in actual materials 
via the f-sum rule \cite{Maldague}.

The kinetic energy is evaluated from the following thermal average,
\begin{eqnarray}
   K  = \frac {1}{Z} \sum_m e^{-\beta E_m} \langle m | {1 \over N_s} \sum_{{\bf k},\sigma} \epsilon_{\bf k}
      c_{{\bf k}\sigma}^{\dagger} c_{{\bf k}\sigma} | m \rangle,
\end{eqnarray}
where $|m \rangle $ is the total set of eigenstates of the system with energies $E_m$.  

(iii) {\em  Double occupancy:}  
It is useful to analyze the number of double occupancy per site in the lattice which reads:
\begin{equation}
d  =\sum_m e^{-\beta E_m} \langle m | {1 \over N_s} \sum_i n_{i\uparrow}n_{i\downarrow} | m \rangle,
\end{equation}
and is different for the different phases analyzed. For example, it is a key quantity in the analysis of the
Mott transition in the half-filled Hubbard model since $d$ is suppressed in the Mott insulator, which allows to determine the 
 critical Coulomb coupling. 
  In model Eq. (\ref{eq:EHM}), $d$ is helpful for characterizing the  
different possible CO states for different $U$ and $V$.

(iv) {\em Specific heat:}  
From the total  energy of the system, $E=\langle H \rangle$,  we can obtain the specific heat by taking
the derivative  with respect to the temperature, $T$:  
\begin{equation}
C_V={\partial  \langle H \rangle  \over  \partial T}.
\end{equation}

Unless otherwise specified, we use units such that $k_B=\hbar=1$. 
The finite temperature method recovers the ground state 
properties by taking the limit: $\beta \rightarrow \infty $. In practice this is achieved for $\beta=50-100$ for the
various $U$ and $V$ explored across the whole phase diagram.
Typically about 30 to 50 terms  
are kept in the evaluation of the statistical sums over the excited
states $|m \rangle$  with corresponding energies  $E_m$.

\section{Phase diagram at T=0}

\label{sec:T0}

\begin{figure}
 \center
   \includegraphics[clip,scale=0.7]{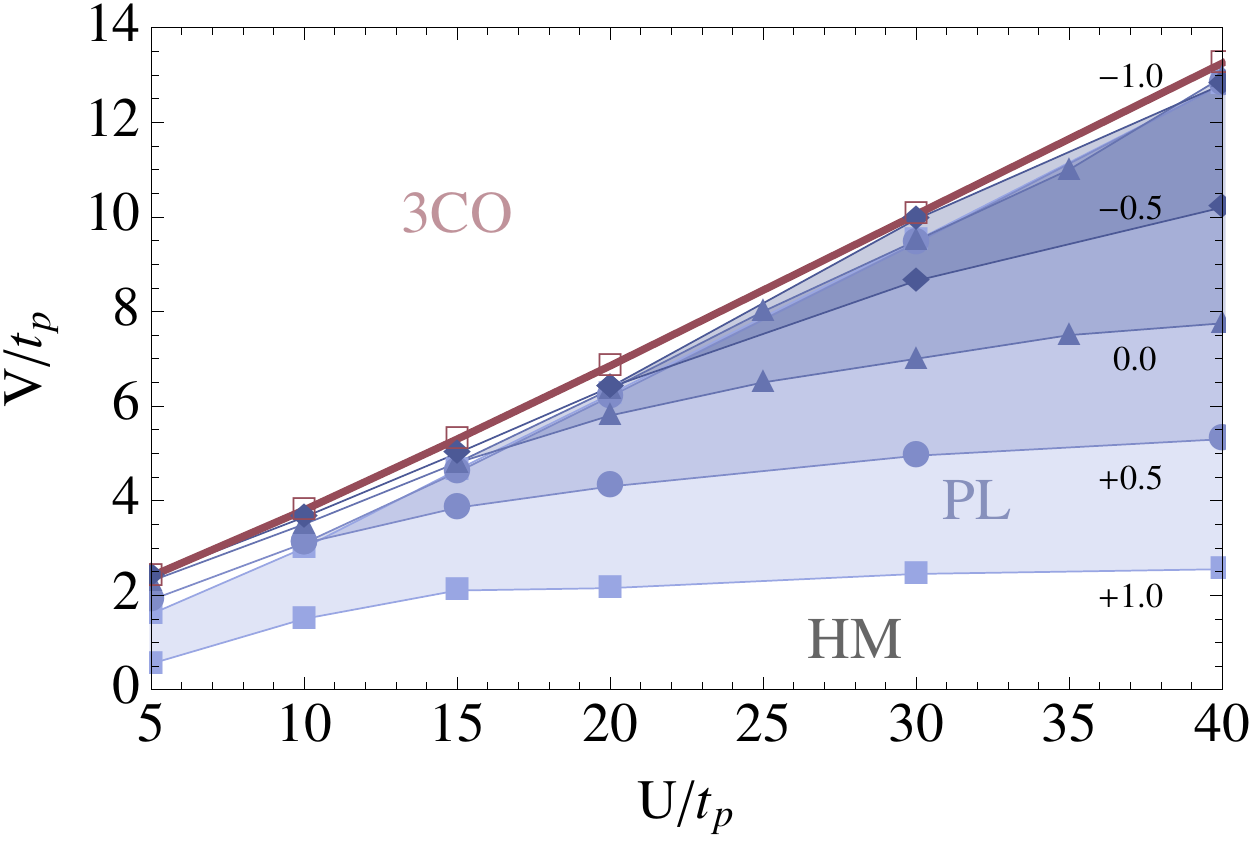}
   \caption{\label{phaseVU_T0}(Color online)
    Phase diagram obtained at $T=0$ from numerical diagonalization of 
    a $N_s=12$ cluster. The different shaded areas correspond to the
    pinball (PL) phase for $t_c/t_p=1,0.5,0,-0.5$ (squares, circles,
    triangles, diamonds respectively, from light to dark blue). 
  The case $t_c/t_p=-1$ has no
    pinball phase in the explored $U$ range. The red line corresponds
    to a direct transition from the homogeneous metal (HM) or pinball liquid (PL) to 
    the threefold charge ordered (3CO) state. }
\end{figure}

The zero temperature phase diagram of the model Eq. (\ref{eq:EHM})
in the $(U,V)$ plane is shown in Fig. \ref{phaseVU_T0}. 
The phase transition lines are determined
using three alternative methods, that all give coincident results: (i) via
the evolution of the charge correlation function $C({\bf Q})$ calculated 
at the three-fold wavevector, Fig. \ref{fig:physical}; (ii)
by  tracking directly the charge ordering patterns that develop
in real space, Fig. \ref{fig:dens}; 
and (iii) by analyzing the {\it fidelity} between
groundstates at different values of the microscopic parameters,
Fig. \ref{fidelity}, as introduced below. 

We are interested here in the charge ordering instabilities 
driven by the inter-site repulsion $V$.
Our numerical results confirm the existence of three
distinct phases: a homogeneous metal (HM) at low $V$, 
a three-fold charge ordered phase (3CO) at large $V$, and an intermediate
``pinball liquid'' (PL) phase emerging at large values of $U$. 
The most striking effect in Fig. \ref{phaseVU_T0} is that the region
of the homogeneous metallic phase is strongly reduced upon increasing
the $t_c/t_p$ ratio, evidently 
due to a corresponding stabilization of the competing pinball liquid phase.  
We note that the homogeneous metal is always the ground state at $V=0$
independently of the strength of the local
repulsion $U$.  This can be rationalized by the fact that 
in the absence of nearest-neighbor interactions, 
at $n=3/2$  
the holes ($n_h=2-n=1/2$) can effectively avoid each other when moving
along the lattice.

\subsection{Characterization of the different phases}

\begin{figure*}
 \center
    \includegraphics[clip=0,scale=0.3]{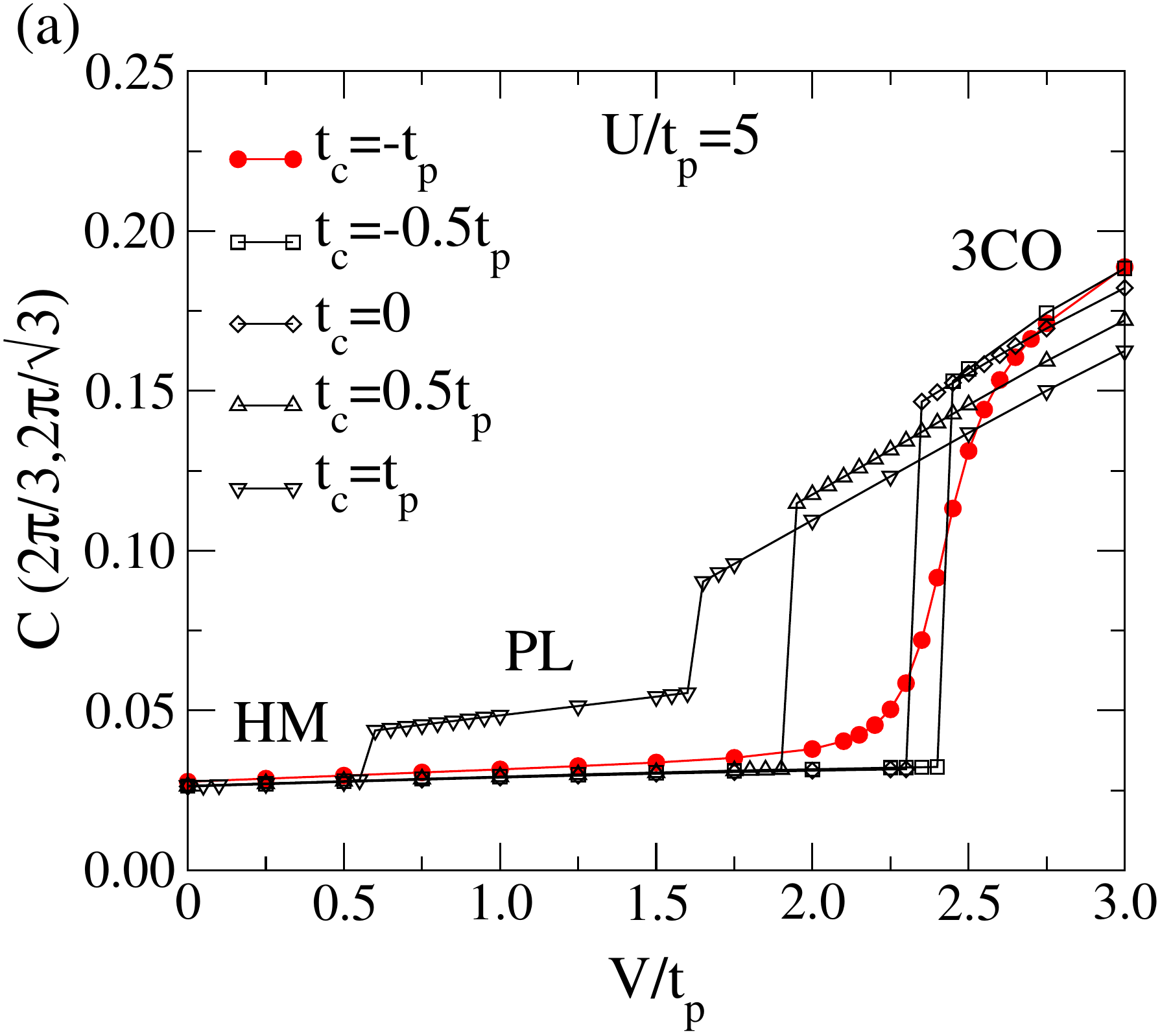}
    \includegraphics[clip=0,scale=0.3]{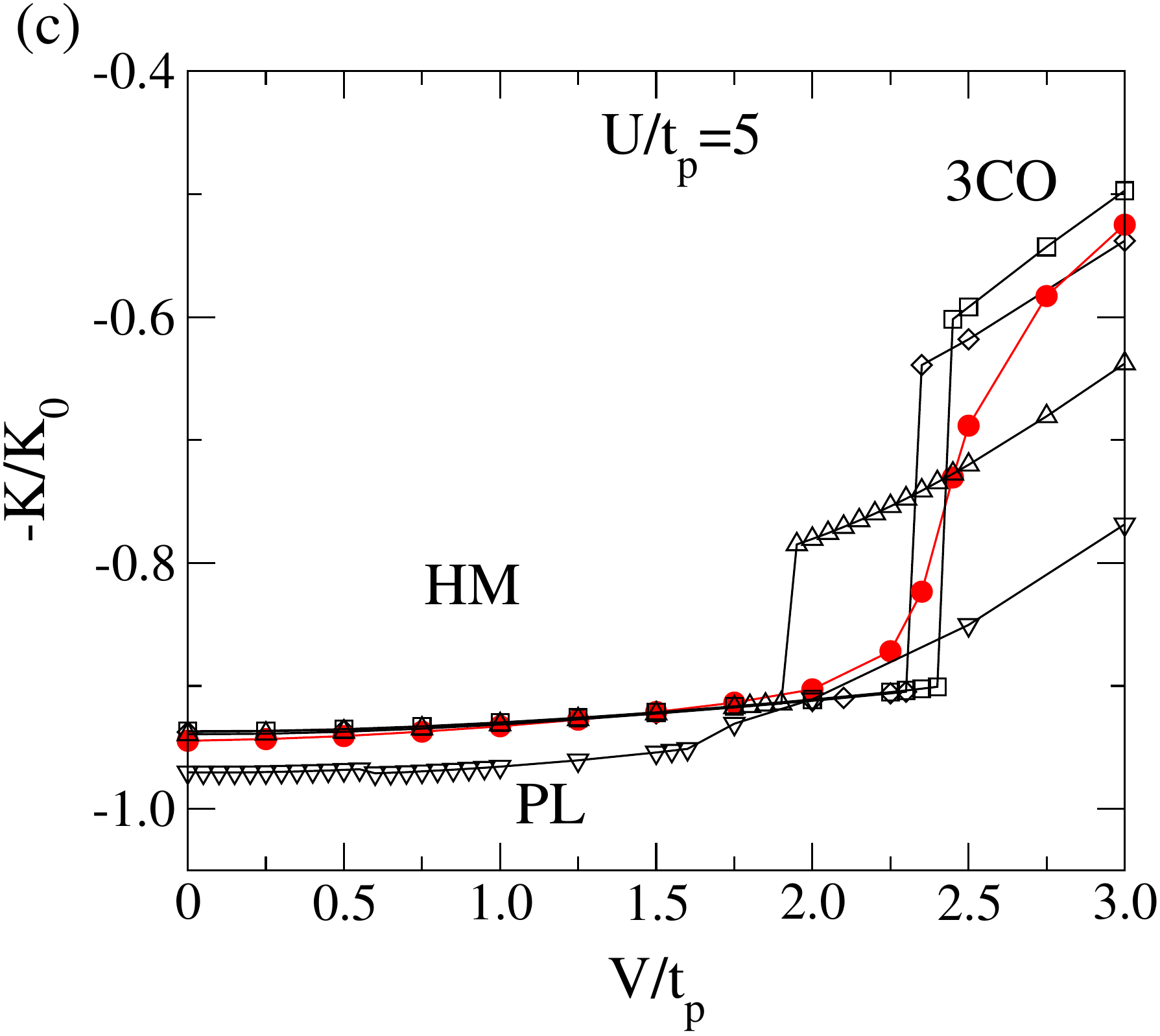}
    \includegraphics[clip=0,scale=0.3]{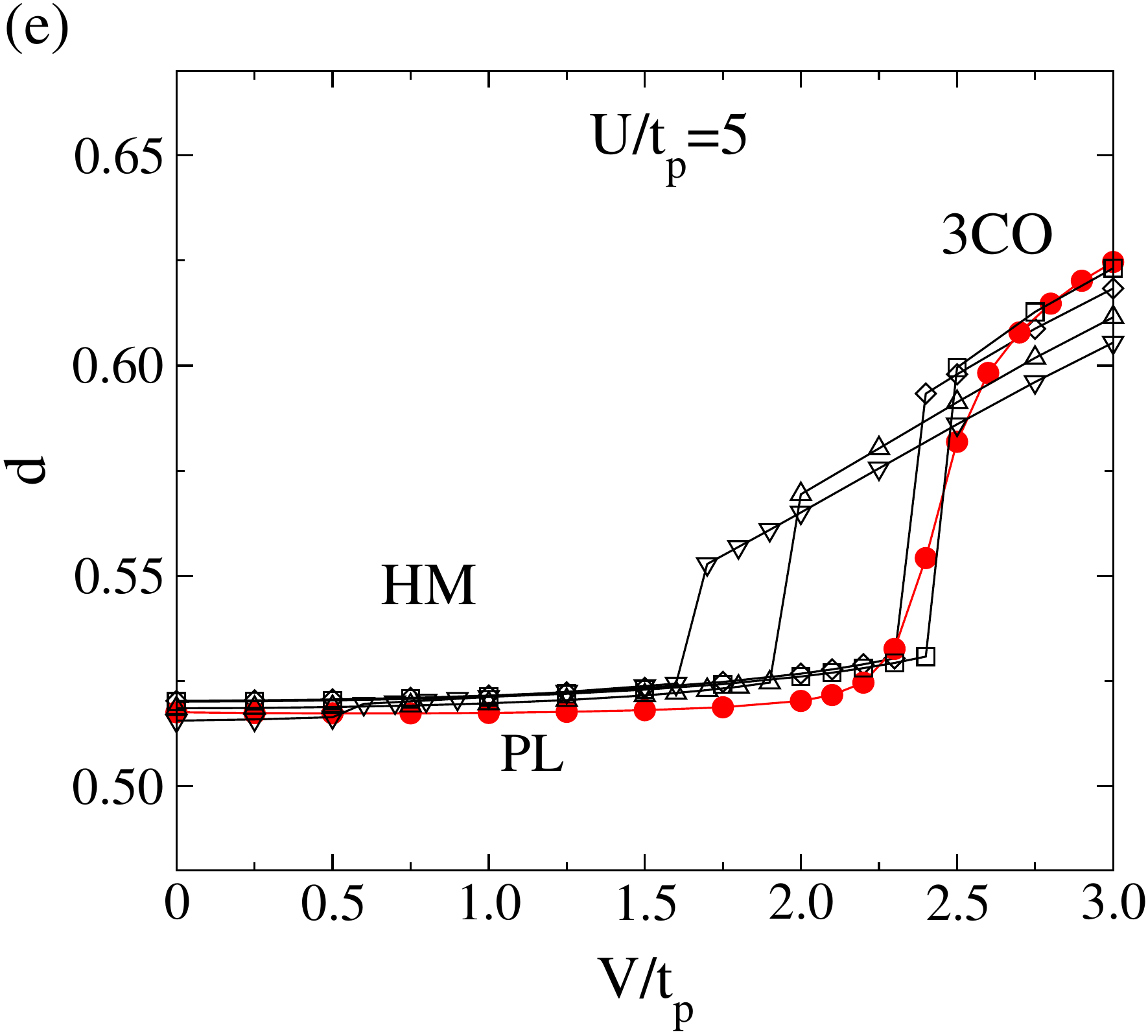}
    \includegraphics[clip=0,scale=0.3]{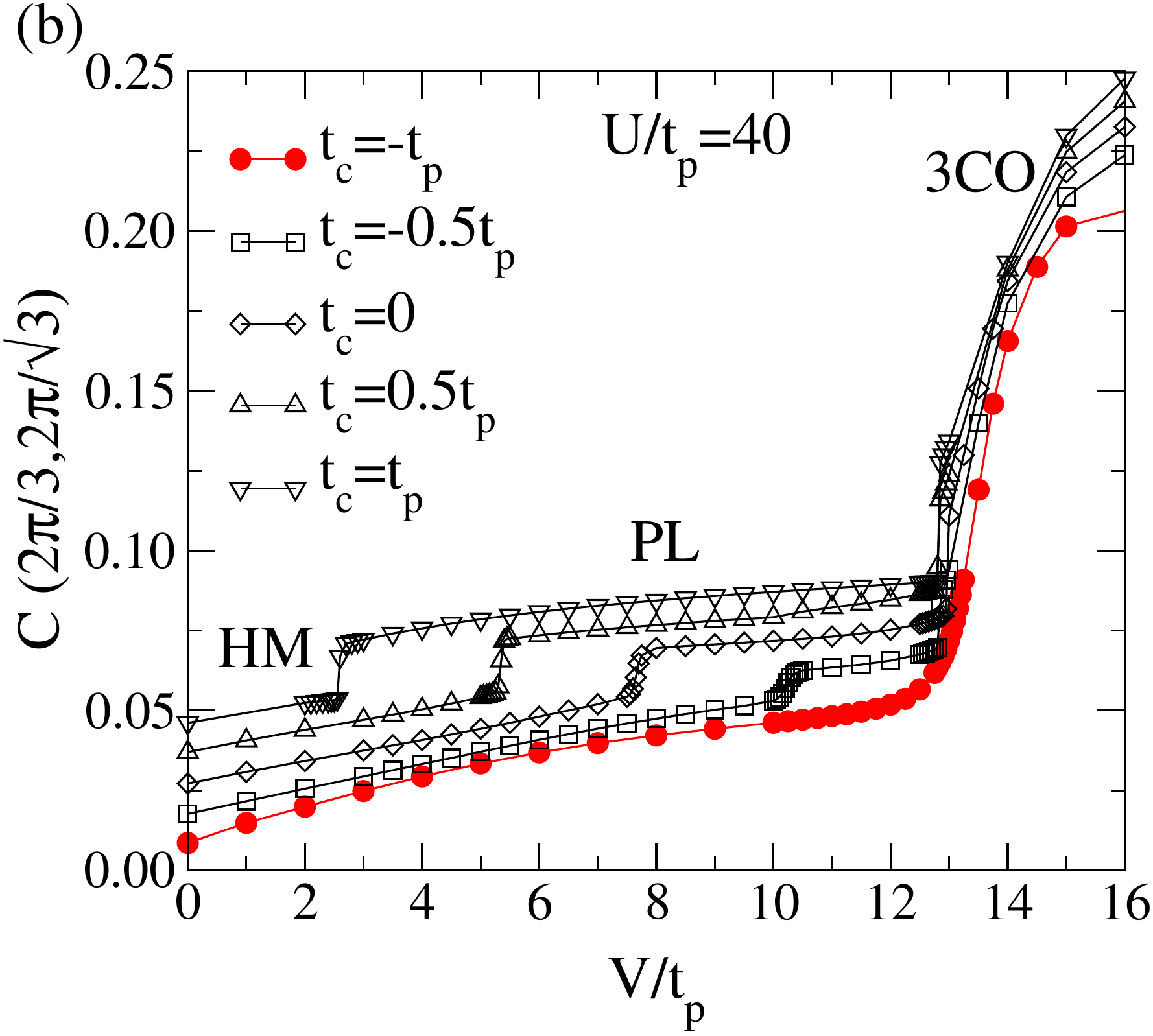}
    \includegraphics[clip=0,scale=0.3]{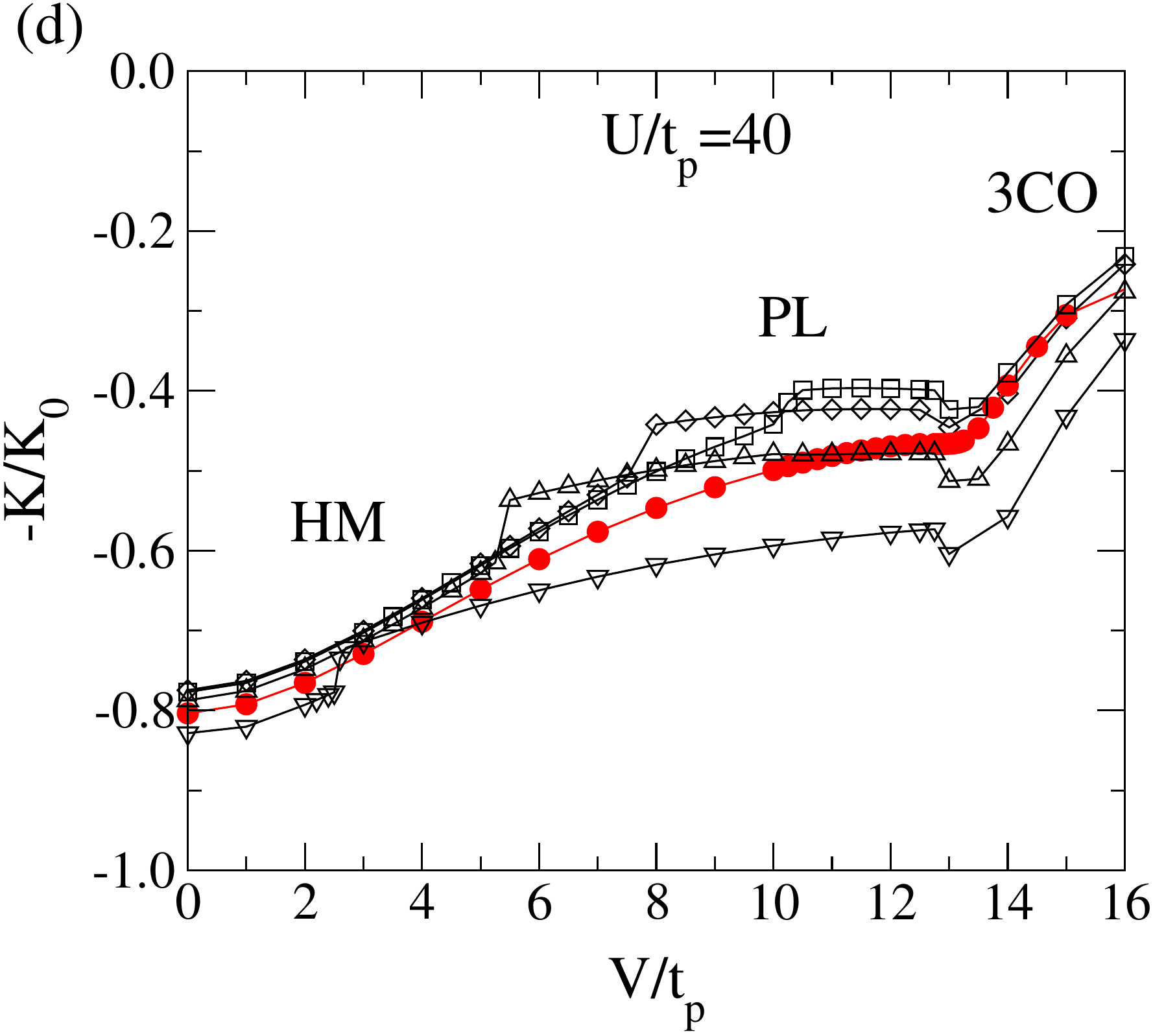}
    \includegraphics[clip=0,scale=0.3]{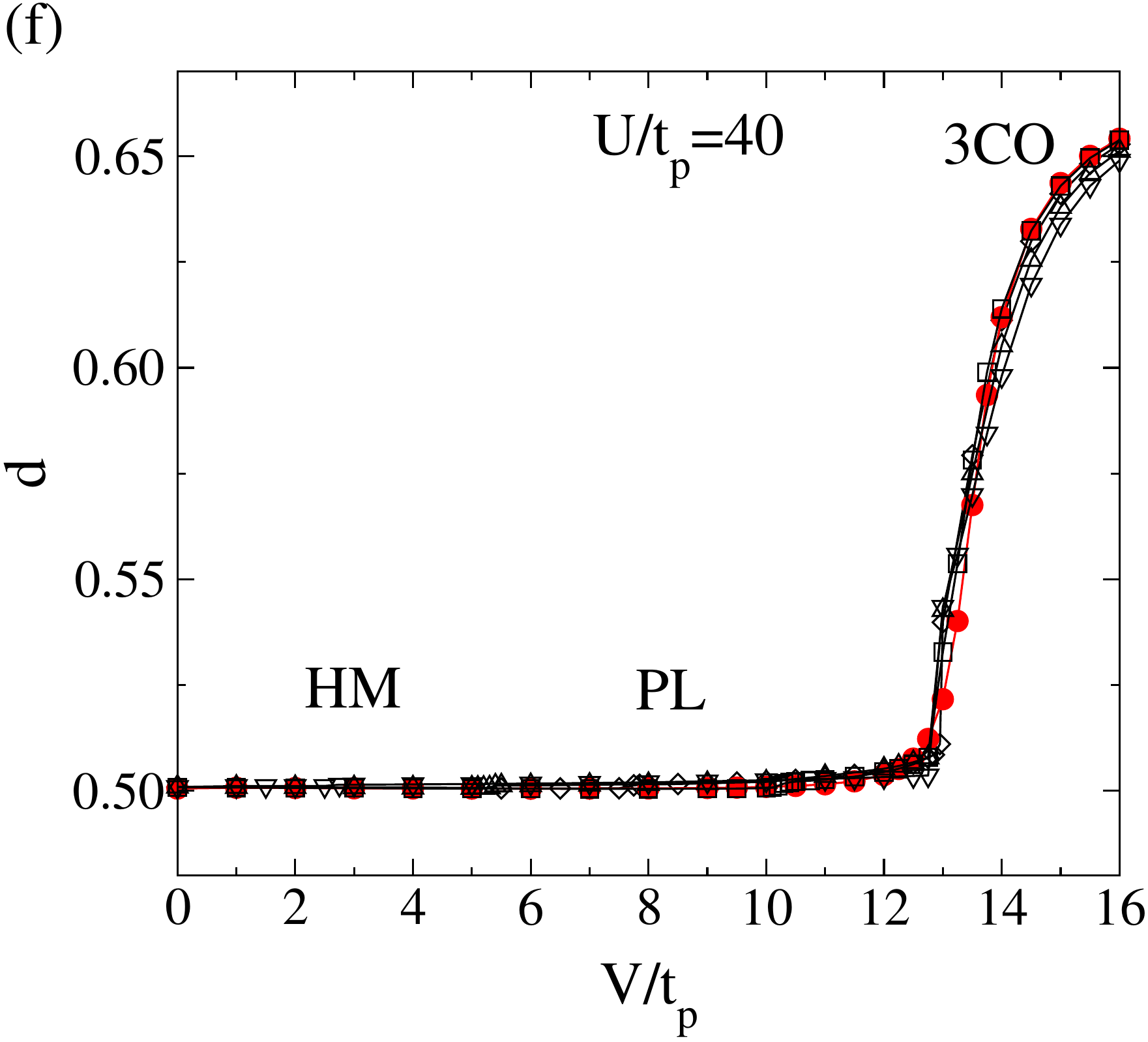}
    \caption{\label{fig:physical}(Color online)
    (a)-(b) Charge correlation function for the threefold
    wavevector ${\bf Q}=(2\pi/3,2\pi/\sqrt{3})$ 
    as a function of $V/t_p$ (a) in the weakly correlated  
    $U=5t_p$  and (b) the strongly correlated limit $U=40t_p$.
    (c)-(d) Average normalized kinetic energy $K
    /K_0$  as a function of $V/t_p$.  (e)-(f)
     Double occupancy probability. The plots in (b) have a vertical offset of  $\pm 0.02$ for $t_c= \pm t_p$ 
     and of $\pm 0.01$ for $t_c \pm 0.5 t_p$ for clarity. }
     \end{figure*}

Fig. \ref{fig:physical} reports the evolution of the charge
correlation function, the kinetic energy and double occupancy
as a function of $V$, along vertical cuts in the phase diagram
corresponding to $U/t_p=5$  and $U/t_p=40$.
Different curves correspond to different values of $t_c/t_p$ (upper and lower panels,
respectively).

\subsubsection{Small $U$: HM to 3CO transition}

At low $U$, the instability towards the threefold charge ordered phase 
 is signaled by a
sharp jump in the correlation function (Fig. \ref{fig:physical}a),
starting from a small constant value in the homogeneous metal.
The locus of the 3CO  transition shows an appreciable dependence on
geometrical frustration: the homogeneous metal is rapidly destabilized
for positive values of $t_c/t_p$, 
in marked contrast with the weaker (and opposite)
variations expected from an RPA analysis valid in the weakly
correlated limit \cite{Kuroki06} (see Appendix \ref{RPA}). 
From the phase diagram of Fig.  \ref{phaseVU_T0} it is quite clear that
this trend is governed by a mechanism that extends from the
strongly correlated limit $U\gg t_p$ down to the lowest values of $U$.  
The emergence of an intermediate plateau 
in the charge correlation function, clearly visible in the data at
 $t_c=t_p$ in Fig. \ref{fig:physical}a, is also 
reminiscent of the situation encountered at $U/t_p=40$ (see below).  
These observations suggest that the presence of 
geometrical frustration, $t_c/t_p>0$,
strongly enhances the role of electronic correlations. 
The pinball phase characteristic of strong $U$ is
stabilized at  $t_c=t_p$ despite a relatively low nominal value $U/t_p=5$. 

We note that the nature of the ordering 
transition changes in the opposite limiting case
$t_c=-t_p$, where the sharp jump in the correlation function is
replaced by a smoother evolution, possibly due to the competition with
an incipient nesting instability (Appendix \ref{RPA}). 

The behavior of the charge correlation function is directly mirrored
in the  other physical quantities 
shown in Figs. \ref{fig:physical}. The kinetic energy
(Fig. \ref{fig:physical}c)  
 jumps at the phase transition from an essentially free-electron
 value, $K/K_0\gtrsim 0.9$, to a value that is reduced by the opening
 of the charge ordering gap. 
At the same time, the double
occupancy (Fig. \ref{fig:physical}e)
undergoes a marked increase  towards the value $d=0.66$ of the fully formed 3CO:
the charge is ordered into three sublattices with
average occupations $n_A=n_B=2$ and $n_C=1/2$ (see Fig.\ref{fig:dens},
large $V$ region), 
so that each of the two charge rich sublattices contributes
$d_A=d_B=1/3$ to the average double occupancy. 
The fact that the double occupancy  in the homogeneous
metal is suppressed from the non-interacting value $d=(n/2)^2= (3/4)^2=0.5625$ 
indicates the presence of moderate electronic correlations.

\subsubsection{Large $U$: HM to PL transition}

A richer situation is found in the large $U$ regime. First of all,
the homogeneous liquid is 
characterized by a total
suppression of double occupancy: introducing the double occupancy of holes 
$d_h=\langle(1-n_{i\uparrow})(1-n_{i\downarrow})\rangle =
1-n+d$ and setting $d_h=0$  
we  obtain $d=n-1=1/2$, which is actually observed in
Fig. \ref{fig:physical}f. Furthermore, the presence of both local and
non-local Coulomb interactions hinders the particle motion, 
resulting in a marked reduction of the kinetic energy 
upon increasing $V$. 
An approximate expression for its $V$-dependence is:
\begin{equation}
K^{(HM)}  \approx (1-AV^2)  K_U^{(HM)}  ,
\label{approx}
\end{equation}
where $K_U^{(HM)}  $ is the value at $V=0$.
This $V^2$ dependence is  consistent with a previous slave-boson
\cite{Seibold} 
calculation of  the metallic phase formed by spinless particles on a
$d$-dimensional hypercubic lattice, which 
is compatible with the data of
Fig. \ref{fig:physical}d.
Because  the ordering instability is pushed to large values of $V$
for negative values of $t_c/t_p$
(cf. Fig. \ref{phaseVU_T0}), the homogeneous metal that is so
revealed  can acquire quite a strongly correlated character,
as testified by a kinetic energy ratio that decreases down to
$K/K_0\simeq 0.5$ before the onset
of charge order.

The numerical
data of Figs. \ref{fig:physical}b and d
show quite clearly that an intermediate phase emerges between the
homogeneous metal and the 3CO, that we associate with the pinball
liquid phase introduced by Hotta and
coworkers\cite{Hotta06,HottaFurukawa06,Miyazaki}. The PL is a
partially ordered phase with a three-sublattice structure, 
in which the carriers of one sublattice (pins) are localized
as a Wigner-crystal  and the remainder (balls) form an essentially
non-interacting liquid within the resulting hexagonal lattice (Fig.1b). 

\begin{figure}
  \centering
  \includegraphics[scale=0.6]{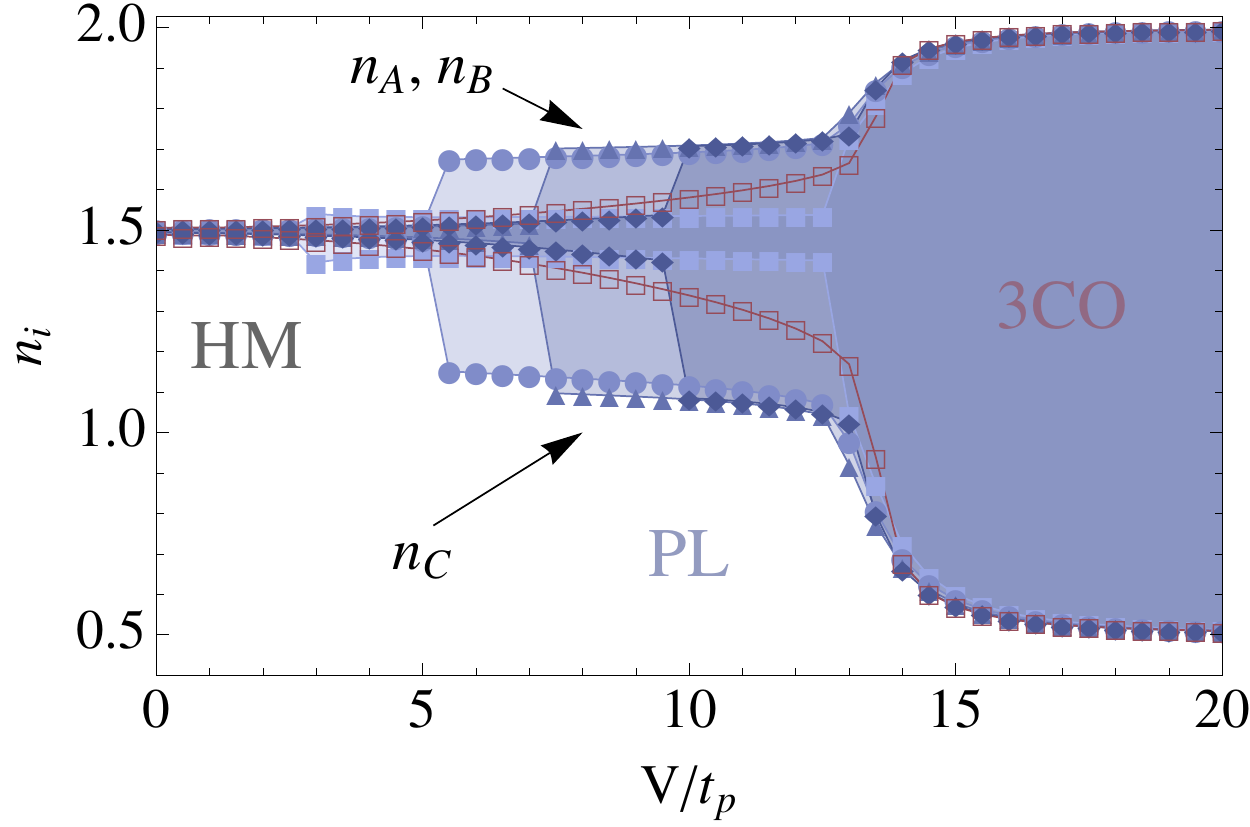}
  \caption{(Color online) Average electron densities in the electron rich (hole poor,
    $n_A=n_B \ge n $, upper
    curves) and electron poor (hole
    rich, $n_C\le n$, lower curves) sublattices, as obtained from
    Lanczos diagonalization at $U/t_p=40$ 
    in the presence of a weak translational symmetry
    breaking potential. We take $\delta=+0.05 t_p$ on the sites of the 
    hole rich sublattice in order to select one realization of the 
    threefold symmetry. Color codes are the same as in
    Fig. \ref{phaseVU_T0} (from light to dark blue:
    $t_c/t_p=1,0.5,0,-0.5$, red: $t_c/t_p=-1$).}
  \label{fig:dens}
\end{figure}

To get a further insight about the different broken symmetry phases, we have calculated the static density profile in the presence of a local 
perturbation breaking all the translations of the lattice, but respecting the $\pi/3$ rotation. In this way, instead of 
obtaining a uniform linear combination of all symmetry related crystal states, the system selects one crystal state 
favored by the perturbation, giving access to a real-space snapshot of the broken-symmetry ground state. 
The basis of the method employed for getting the real space snapshot of the local densities is the following: 
we add a local potential on certain sites related by rotations in such a way that only translations are broken. 
The value of the local defects is few percents of the hopping term. In the Hamiltonian, we simply add the perturbation term:
$\delta \sum_{i} \left( n_{i\uparrow} + n_{i\downarrow} \right)$.
This method has previously been used for distinguishing the exact nature of two phases breaking the translational and 
the rotational symmetries differently \cite{Ralko}. Since the additional term is kept very small, it only corresponds to a perturbation and do not change the main property of the ground state. In this paper, we do want to keep the rotational symmetry safe, we hence chose 4 sites, in the 12-site cluster, in such a way that the system is still rotationally invariant.
The perturbation 
has to be lower than the typical energy scale and we put $\delta = + 0.05 t_p$ on the sites of the hole rich sublattice. 
Results are depicted in Fig. \ref{fig:dens}. Expressing the charge densities in
terms of holes, $n_h=2-n=1/2$ and starting from the 3CO phase ($n_{h,A}=n_{h,B}=0$ and $n_{h,C}=3/2$,
Fig. \ref{fig:dens}), by progressively increasing the local repulsion $U$, it becomes
energetically unfavorable to acommodate more than one hole per 
molecule. Part of the hole density will then tend to spill out of the
hole-rich sites in order to prevent double occupancy, resulting in
$n_{h,C}=1$.  The average charge density in the
three sublattices, $n_A=n_B=7/4$, $n_C=1$, illustrated in
Fig. \ref{fig:dens}  indeed corresponds to a scenario where one sublattice is occupied by localized
holes ($n_{h,C}=1$ leading to $n_C=2-n_{h,C}=1$)
with the remaining particles equally spread in the interstitial
sites ($n_{h,A}=n_{h,B}=n_{ball}=1/4$ hole per site, leading to  
$n_A=n_B=2-n_{ball}=7/4$). 
The critical value for the transition from 3CO to PL can be
readily estimated to be $U\simeq 3V$ from electrostatic considerations alone
(see Appendix \ref{Hartree}), and is therefore independent of $t_c/t_p$.

Notably, in the PL phase the physical quantities whose $V$ dependence is
depicted in Fig. \ref{fig:physical} form well defined plateaus,
suggesting that they are locked 
as a consequence of the spontaneous separation between
localized and itinerant charges. For example,
the charge correlation function tends to the value $C({\bf
Q})=n^2/3\simeq 0.08$ instead of the full $C({\bf
Q})=n^2$ obtained at complete ordering in the large $V$ limit, 
corresponding to the fact that only $1/3$ of the particles participate
to the ordering phenomenon.

From our numerical results in Fig. \ref{fig:physical}d,   
the kinetic energy of both the homogeneous metal at $V<V_{PL}$ and 
the 3CO phase at $V>V_{3CO}$ is found to
depend only weakly on the degree of geometrical frustration $t_c/t_p$.
On the other hand, the kinetic energy forms a plateau within the PL phase,
at a  value which is strongly dependent on $t_c/t_p$: 
the absolute value of the (negative) kinetic energy in the PL phase is lowest 
at $t_c=-0.5t_p$ 
and its magnitude steadily increases  
as the frustration ratio is increased to  $t_c=t_p$. 
Our data therefore suggest that 
it is this gain in kinetic energy for positive values of 
the geometrical frustration ratio that is responsible for the
stabilization of the pinball liquid phase against the homogeneous
metal observed in Fig. \ref{phaseVU_T0}.   
\footnote{A word of caution is needed here since due to accidental degeneracies of the
noninteracting kinetic energy, $K_0$ is independent of $t_c/t_p$ in the
small $N_c=12$ cluster used which coincides with the $t_c=0$ kinetic energy of
the extended tight-binding model of the lattice. At $U>5t$ degeneracies are
split and dependence on $t_c/t_p$ is recovered as expected. }
We note that a kinetic energy driven
mechanism for the PL transition was also pointed out in Ref. \onlinecite{Miyazaki}. 
Since the effective filling associated with the itinerant balls on the hexagonal lattice is
only $n_{ball}=1/8$,  Coulomb interaction effects are small.
Indeed, we have checked that the 
kinetic energy of the balls coincides with the kinetic energy of the corresponding 
non-interacting tight-binding model on a hexagonal lattice, at filling $n=n_{ball}$,
as a function of the ratio $t_c/t_p$. 

\subsection{Fidelity analysis}

\begin{figure*}[t!]
\includegraphics[width=0.98\textwidth,clip]{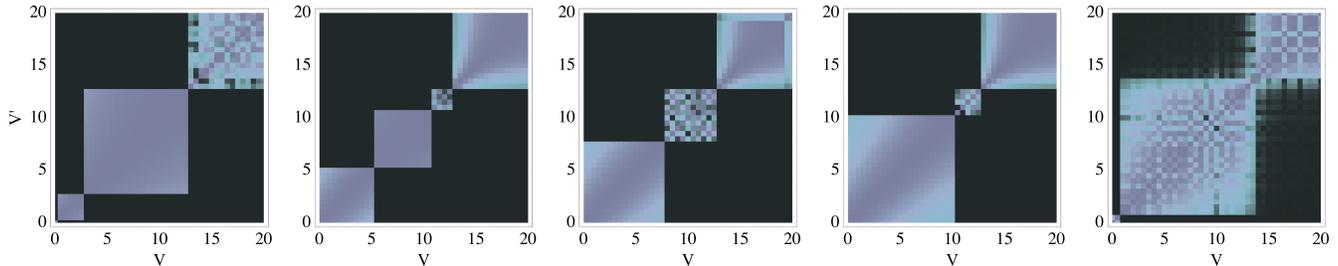}\\
\caption{(Color online).  
Fidelity $F_{V,V^\prime} = | \langle \psi_0 (V^\prime)  
| \psi_0 (V)  \rangle |$ for
various values of $t_c/t_p$, from  left to right $1.0 , 0.5, 0.0, -0.5$ and
$-1.0$ and $U=40|t_p|$. There is a perfect agreement with the phase transitions obtained by
standard quantities (order paramaters, kinetic energies,
suceptibilities). 
The case of zero overlap is represented in black, while the blue-violet (diagonal squares) represents
overlap 1 ($V$=$V^\prime$). The non-uniform colored zones stand for fluctuating
ground states possibly related to a strong degeneracy.
\label{fidelity} }
\end{figure*}

In order to characterize the quantum phase transitions  of our system,
the simple yet powerful concept of {\it fidelity} \cite{uhlmann} is
considered here (Fig. \ref{fidelity}).
Initially introduced in quantum information, the fidelity has revealed
successful in determining superfluid-insulator transitions of the Hubbard
model\cite{nielsen,campos,buonsante}.  The idea behind the fidelity is very
simple; it consists of computing overlaps of ground states (GS) 
$| \psi_0  \rangle$ at different
values of the microscopic parameters. 
At a QCP, even the smallest change of the parameters can have
dramatic effects in some of the observables.  This is encoded in the GS
properties, hence, the overlap is expected to strongly react and indicate the
locations of the phase transitions.  We define the fidelity $F$ as:
\begin{eqnarray} F_{V,V^\prime} = | \langle \psi_0 (V^\prime)  | \psi_0 (V)
  \rangle |. 
\end{eqnarray} 
Obviously, for $V=V^\prime$, 
the fidelity should be $F_{V,V} = 1$. Typical
results are depicted in Fig. \ref{fidelity} at $U=40t_p$, and for
five values of $t_c/t_p = \pm 1.0 , \pm 0.5$ and $0.0$.

It is surprising to see how the fidelity is indeed able to pinpoint the phase
transitions. For each of the values of $t_c/t_p$, 
we exactly recover the transitions obtained by more standard methods
in the preceding Sections.  More information is available, however. 
First, there is
always some 
region where $F$ is strongly fluctuating, even though well delimited
in the $(V,V^\prime)$ 
plane (non uniform colored zones). These fluctuations can be
due to a large degeneracy of the GS hence corresponding to a same order but
with destructive interferences. The second important information is
visible in the case $t_c/t_p=0.5$, where the fidelity  
indicates the existence of two distinct
phases in the pinball region [namely $ 5.20(2)
\leq V/t_p \leq 11.00(2)$ and  $ 11.00(2) \leq V/t_p \leq 13.00(2)$], 
suggesting a possible ordering of the mobile charges. 
In fact, a slight
change in the kinetic energy (Fig. \ref{fig:physical}d) appears at this transition,
but the other quantities seem to be insensitive to it. 
The fact that the transition within the pinball phase is
not detected by the charge correlation function
(Fig. \ref{fig:physical}b) nor by the average densities
(Fig. \ref{fig:dens}) tends to show that their properties
remain extremely close. Nevertheless, the fidelity $F$ allows a
precise determination of the sub-phases.

\subsection{Spinless model}

\label{sec:spinless}

We now consider the spinless version of the 
model Eq. (\ref{eq:EHM}), which has been discussed extensively 
in the literature \cite{Hotta06,HottaFurukawa06,Miyazaki,Nishimoto09}.
Since the spinless model only contains charge degrees of freedom, 
by comparing it with the spinful model we can
obtain useful information on the relative role played by charge
fluctuations as compared to 
the spin fluctuations.  

The spinless model reads:
\begin{eqnarray}
  \label{eq:spinless}
  H &=& t_p \sum_{\langle ij \rangle _p} \left( h_{i}^{\dagger} h_{j}+H.c. \right)
  +t_c \sum_{\langle ij \rangle _c} \left( h_{i}^{\dagger} h_{j}+H.c. \right)
  \nonumber \\
  &+& V_p \sum_{\langle ij \rangle _p} n_i n_j + V_c \sum_{\langle ij \rangle _c} n_i n_j,
\end{eqnarray}
where we have  changed the sign of the hopping integrals to 
deal explicitly with holes. As in the preceding Sections 
we consider the case $V_c=V_p=V$ for different values of the $t_c/t_p$ ratio.  
Importantly, for spinless particles  
the physical situation of one hole per two sites implies a half-filled
band, which gives rise to
a spurious particle-hole
invariance that is absent in the spinful case at $3/4$-filling.
Therefore, the thermodynamic
properties as well as the phase transition lines 
become invariant under a change of sign of $t_c/t_p$.
\begin{figure}
 \center
  \includegraphics[scale=0.4]{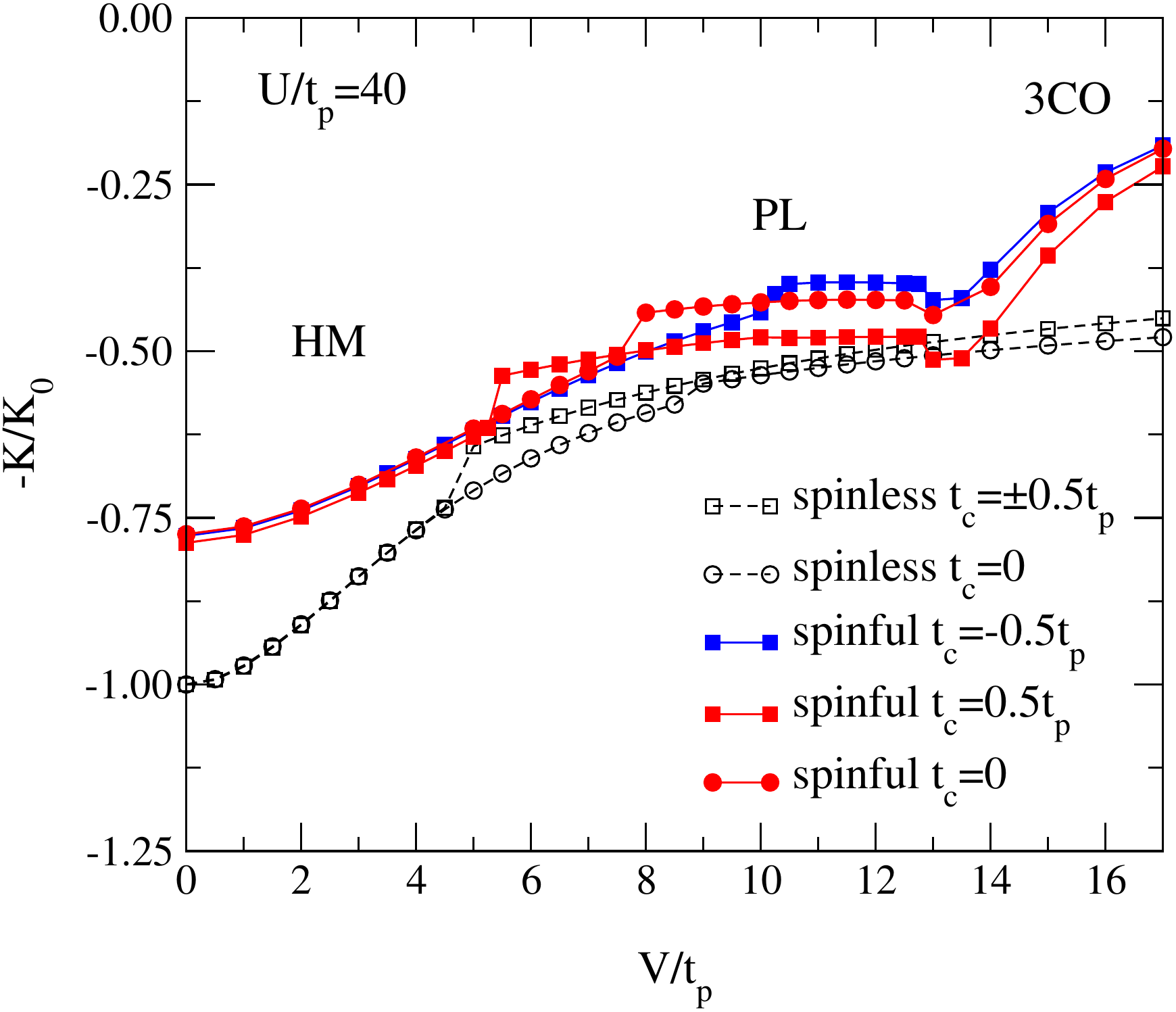}
  \caption{\label{fig:spinless} (Color online) Dependence of the kinetic energy on $V$ for the spinless and spinful extended Hubbard models.
  Different $t_c/t_p$ ratios are compared in the two situations. As expected, the properties of the spinless model
  are independent of the sign of $t_c/t_p$, in contrast to the spinful model. }
\end{figure}

In Fig. \ref{fig:spinless} 
we show the $V$-dependence of the kinetic energy  for different
$t_c/t_p$ ratios compared with the spinful model. As expected the figure shows
how the kinetic energy in the
spinless model does not depend on the sign of $t_c/t_p$ (black curves).
The spinless model is able to recover the
qualitative trends of the full model  for  $t_c/t_p\ge 0$: 
the critical values at which the transition from the HM to
the PL occurs are respectively $V_{PL}\simeq 4.5 t_p$ and $8.5t_p$
for $t_c/t_p=0.5$ and $0$ in the spinless case, to be compared with 
$V_{PL}\simeq 5.5 t_p$ and $7.5t_p$ in the spinful case at $U=40t_p$
(Fig. \ref{fig:spinless}).
However, because of the artificial particle-hole symmetry,
 for $t_c/t_p<0$ the locus of the transition is completely
 inconsistent with the spinful case, 
and the stability of the PL is widely overestimated. 

We see from  Fig. \ref{fig:spinless} that the effect of non-local
interactions $V$ on the 
renormalization of the kinetic energy
 in the homogeneous metal
is very similar  in the spinful model at large $U$  and in the
spinless model, both being compatible with the quadratic
$V$-dependence discussed above. 
We therefore conclude that the different
behaviors observed in the two models at $t_c/t_p>0$ 
are a direct consequence of the different
kinetic energies at the {\it non-interacting} level, that results from the
spurious particle-hole symmetry acquired by the spinless version.

The results presented here indicate that the charge rather than spin
correlations dominate the renormalization  
effects on metallic properties approaching the charge order transition. 
However the spin multiplicity enters (indirectly) via the geometrical
frustration, that is not treated correctly in the spinless model.
A realistic spinful calculation therefore appears to be necessary to
properly address the physics of $\theta$-(ET)$_2$X salts, where $t_c/t_p$ is a
key parameter in determining the experimental phase
diagram\cite{Mori,Kuroki09}.

\section{Correlated metal at finite temperatures}

\label{sec:finiteT}
Here we analyze the properties of the homogeneous metallic phase at finite
temperatures close to the QCP. 
A temperature scale emerges, that we denote $T^*$, above which 
the kinetic energy departs from Fermi liquid behavior and
the specific heat coefficient goes through a maximum.
We interpret $T^*$  as a {\it renormalized Fermi temperature}, that
generally drops to zero at the approach of the QCP. 
Such behavior is typically found for  geometrical frustration $t_c \ne
\pm t_p$. 
On the contrary, in cases in which there are competing Fermi surface instabilities,
especially in the perfectly nested case $t_c=-t_p$, the $T^*$
phenomenon is much weaker, and hardly affects the properties of the
electron liquid. 
The data in that case are compatible with 
a Fermi temperature that remains finite right close to the QCP,
possibly indicating that a first order transition may 
be occurring.    

In the following paragraphs we 
focus specifically on the realistic value $U=15t_p$, 
but the qualitative features presented here are unchanged for different large values of $U$.

\subsection{Non-Fermi liquid behavior close to CO}

\begin{figure}
 \center
     \includegraphics[clip,scale=0.23]{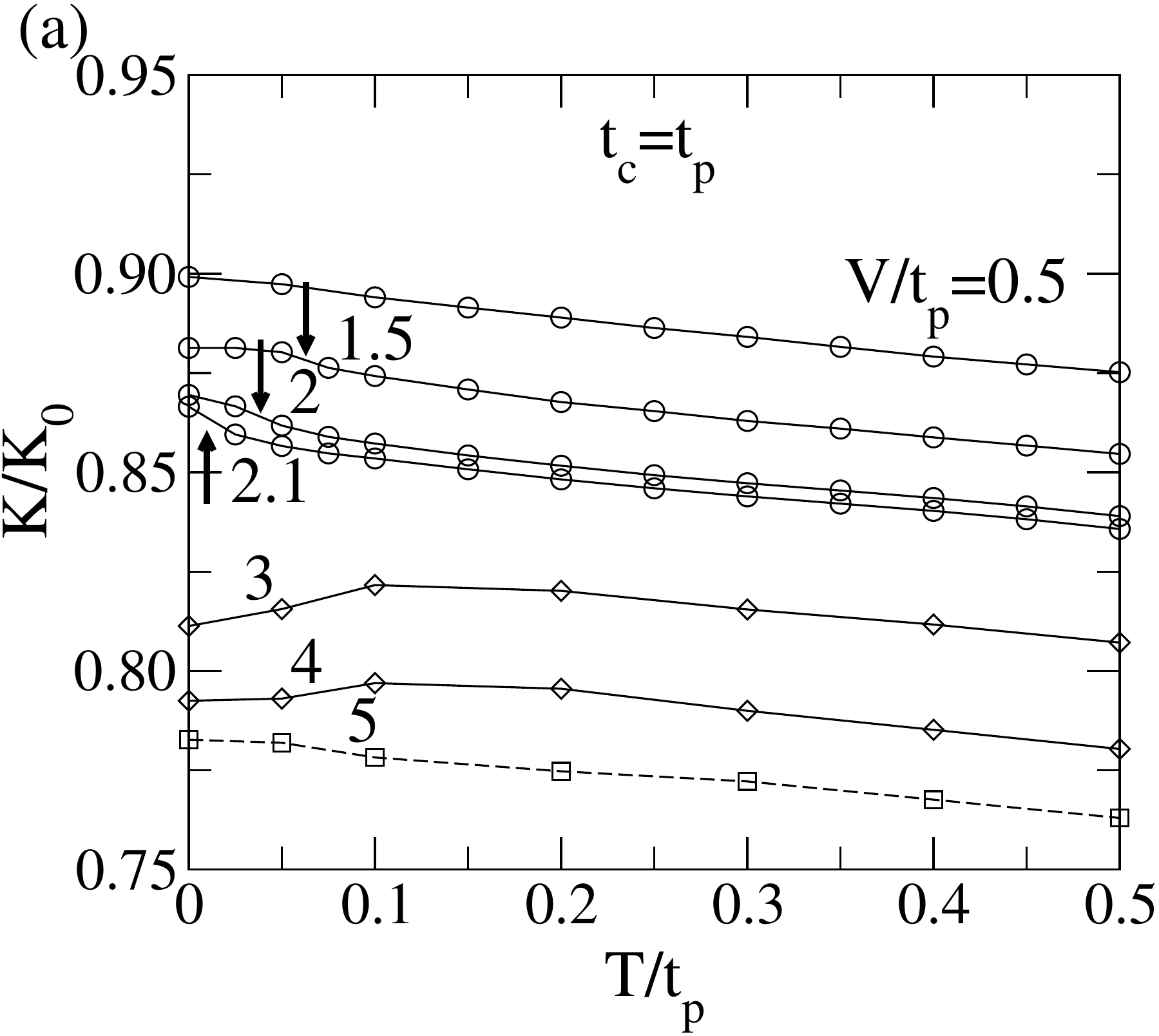}
     \includegraphics[clip,scale=0.23]{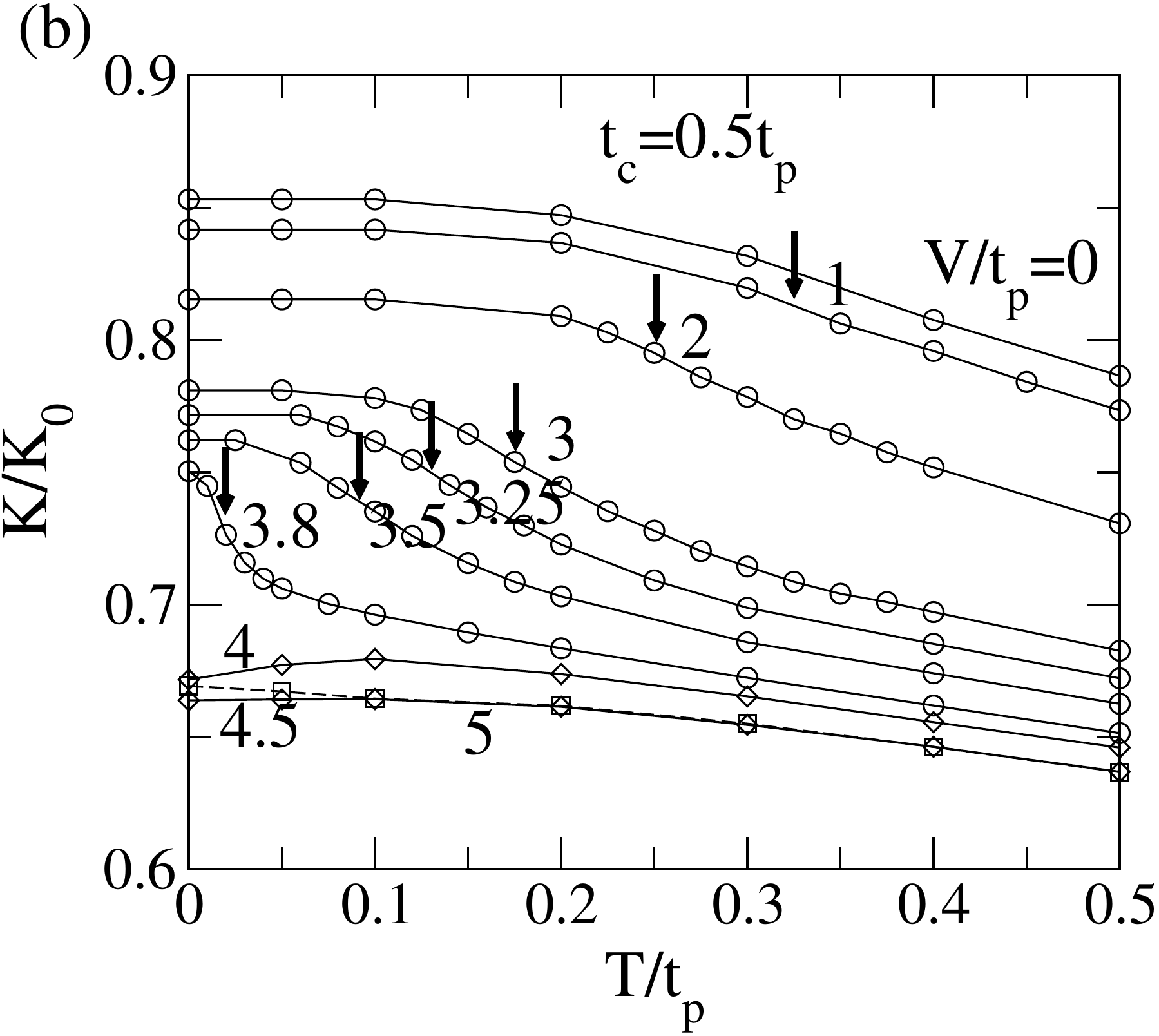}
     \includegraphics[clip,scale=0.23]{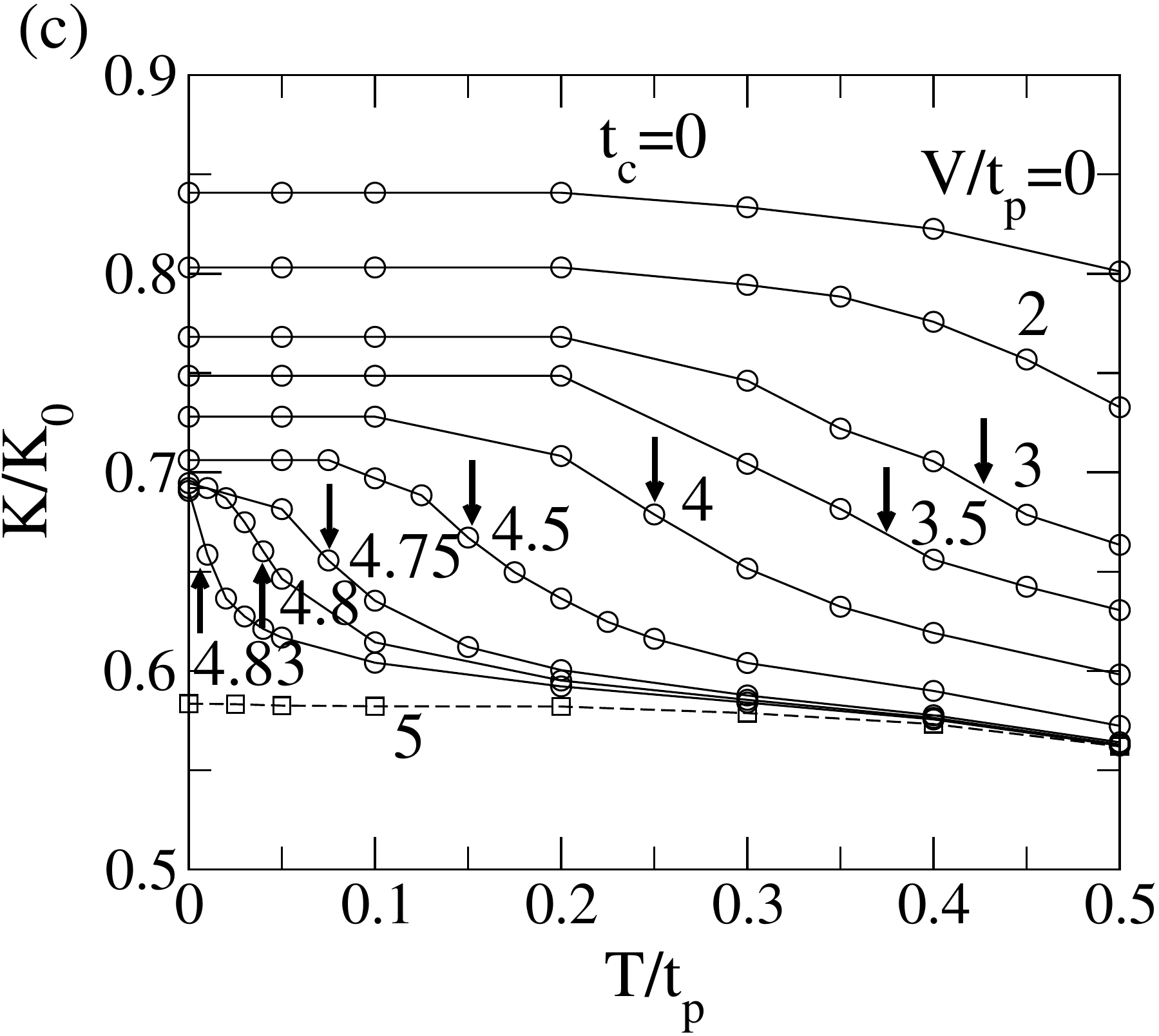}
     \includegraphics[clip,scale=0.23]{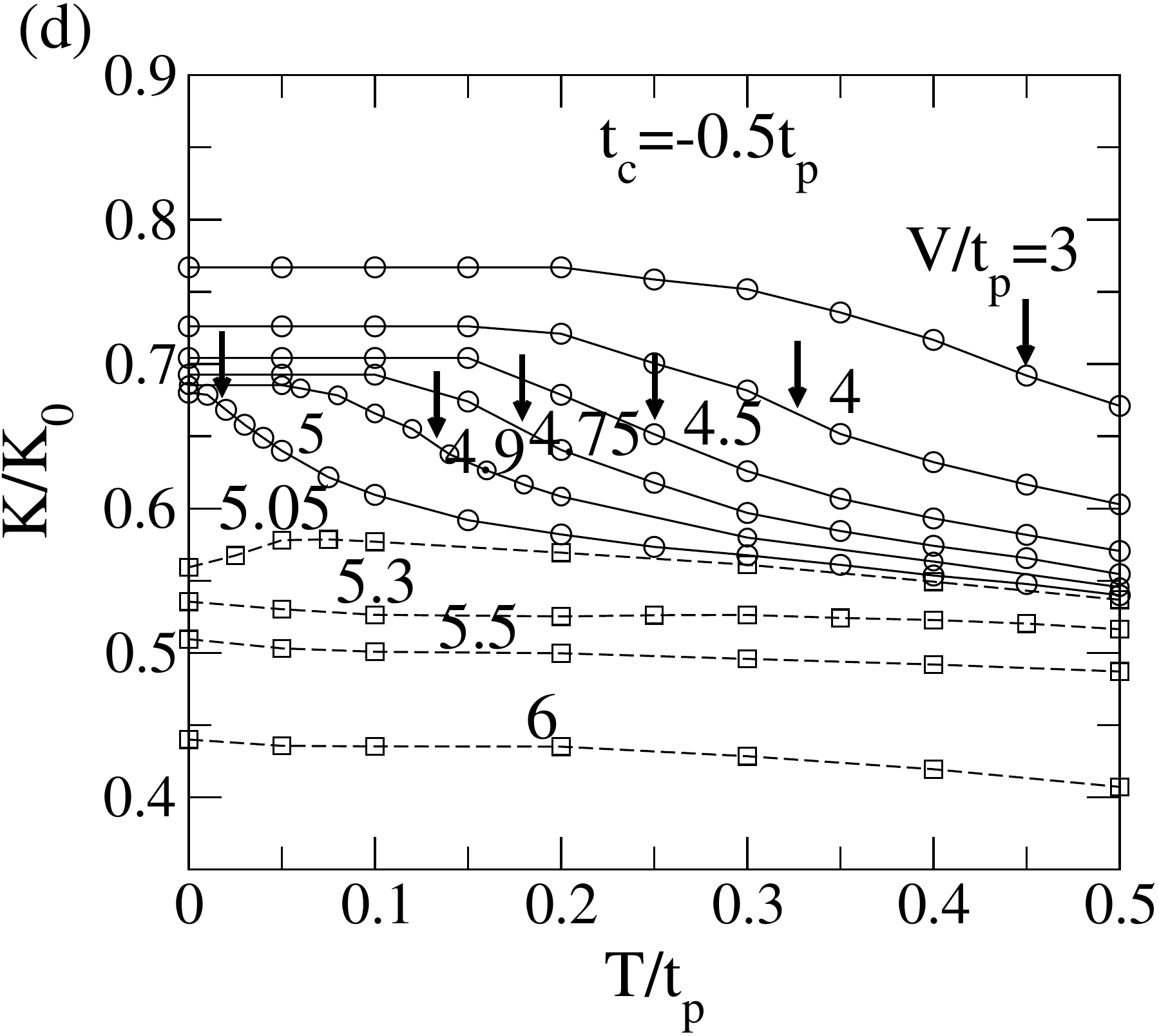}
     \includegraphics[clip,scale=0.23]{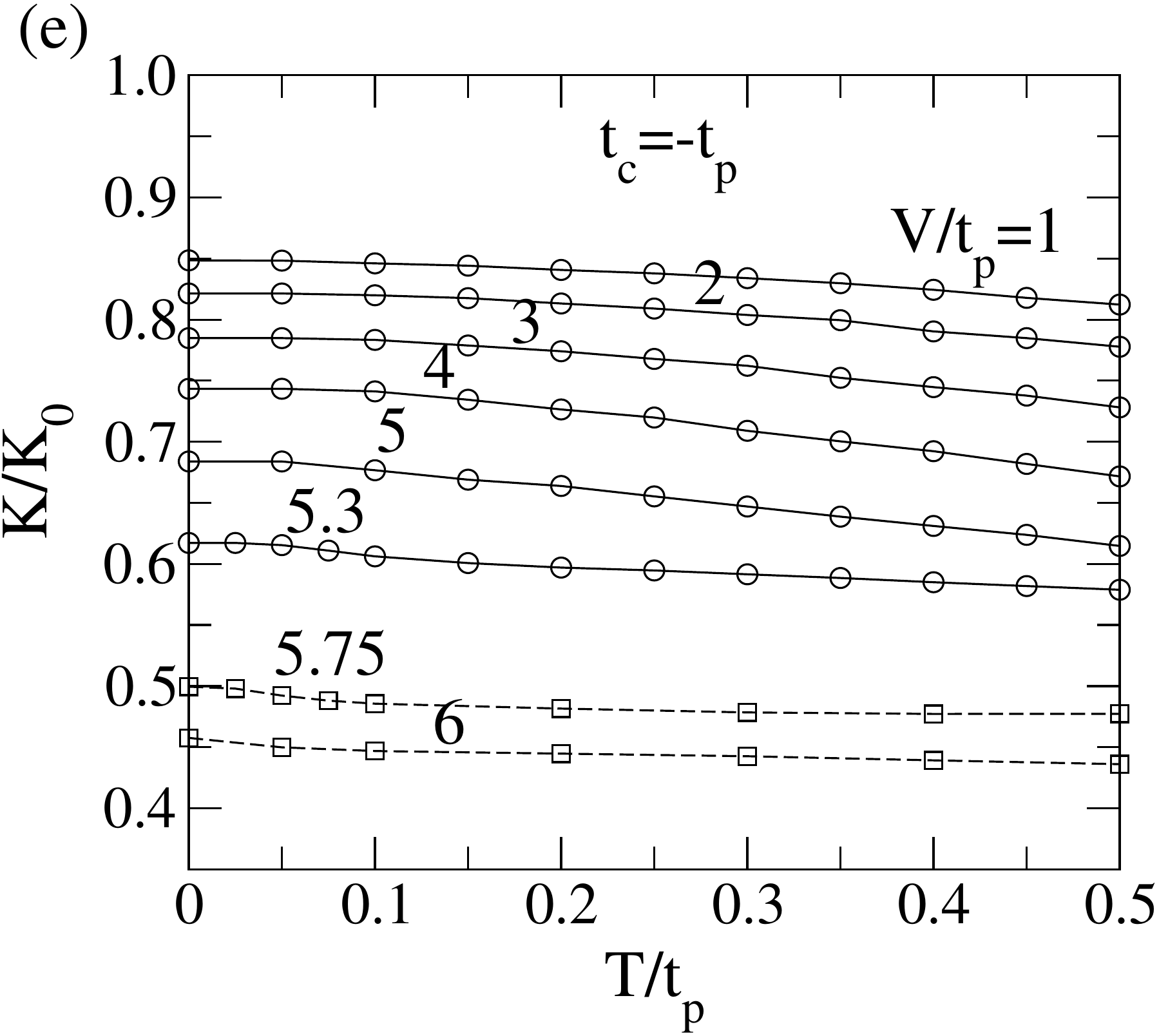}
     \caption{\label{kineticU15V_T}
     Temperature dependence of the 
     total kinetic energy at $U=15t_p$ normalized to the non-interacting
     value $K_0$, for 
     $t_c/t_p=-1,-0.5, 0, 0.5$ and $1$. 
     Arrows correspond to inflection points of the curves, indicating
     the temperature scale $T^*$ corresponding to the breakdown of the
     quasiparticles. 
     Open circles  correspond to the HM phase, diamonds to the PL
     (full lines) and
     squares (dashed lines) to the 3CO phase.}
\end{figure}

\subsubsection{Kinetic energy}
The temperature dependence of the average kinetic energy $K$
normalized to the non-interacting value $K_0$ ($U=V=0$) is  
shown in Fig.~\ref{kineticU15V_T} for $t_c/t_p=1,0.5, 0, -0.5,
-1$. The different curves in each panel correspond to
different values of the intersite Coulomb 
repulsion, $V$, across the charge ordering transitions. 
The  ratios $t_c/t_p=0.5,0,-0.5$ (Fig.~\ref{kineticU15V_T}b, c and d)
all show marked 
departures from the quadratic temperature dependence $K=K_{T=0}-B T^2$
characteristic of conventional metals,
occurring in the HM phase above a certain temperature (the approximate
locus of the inflection points is indicated by arrows). 
We denote it as $T^*$ and take it
as an estimate of  the renormalized Fermi temperature, governing 
a crossover to non-Fermi liquid behavior.  
Clearly,  $T^*$ is progressively reduced upon
approaching the charge ordering transition  and 
vanishes at the critical point.
In this respect, our data in the experimentally relevant case  $U=15t_p$ 
do not show qualitative differences between the transition to the 
 threefold charge order obtained for $t_c/t_p=-0.5;0$ 
 and that to the pinball phase for $t_c/t_p=0.5$ (cf. Fig. \ref{phaseVU_T0}):
in both cases the 
temperature scale $T^*$ appears to be entirely controlled by the
approach to the zero-temperature ordering 
transition, indicating the possibility of quantum
critical behavior at finite temperatures around the zero-temperature
phase transition.

In the cases in which $t_c=\pm t_p$, the emergence of a temperature scale $T^*$
is much less clear [Fig.~\ref{kineticU15V_T} (a) and (e)]
as the temperature dependence of the
kinetic energy is smooth within the whole homogeneous metallic 
phase except very close to the transition.
Our RPA analysis presented in Appendix \ref{RPA} shows that
in these cases there is a competing CDW instability, that
could indeed be  masking the  quantum critical behavior associated with the 3CO transition.
For $t_c/t_p=-1$ the CDW is driven by the perfect nesting of the Fermi surface whereas for 
$t_c/t_p=1$ a mixed CDW/CO phase induced by both nesting tendencies and strong Coulomb
repulsion coexist. 

The perfect nesting of the Fermi surface for $t_c/t_p=-1$
occurs at the wavevectors ${\bf Q}_F=(\pm \pi,\pm \pi/\sqrt{3})$ and $(0,\pm
2\pi/\sqrt{3})$. Such nesting instability is dominant at weak $U$, where it results in a
striped charge modulation that dominates over the threefold charge order
discussed above (this striped order is analogous to the checkerboard
pattern obtained in the square lattice).  Our data suggests that
even in the presence of a sizable local Coulomb repulsion,  $U=15t_p$,
that prevents  the stabilization of such stripe order, 
an incipient nesting instability is strong enough to destroy 
the quantum criticality around the QCP.  This conclusion is based on the observation that
there is no clear signature of the vanishing low temperature scale, $T^*$, at the QCP 
and there is no clear evidence of the 'bad'  metallic behavior found for other $t_c/t_p$ ratios.
However, there is a weak $T$-dependence of the kinetic energy that  vanishes at a critical value 
$V_c=6t_p$ as can 
be observed from the data of Fig.  \ref{kineticU15V_T} (e). Such critical value, $V_c$, is found to be 
consistent with the critical value obtained from the 3CO charge correlations calculated below.

\subsubsection{Charge correlations}

\begin{figure}
 \center
     \includegraphics[clip,scale=0.23]{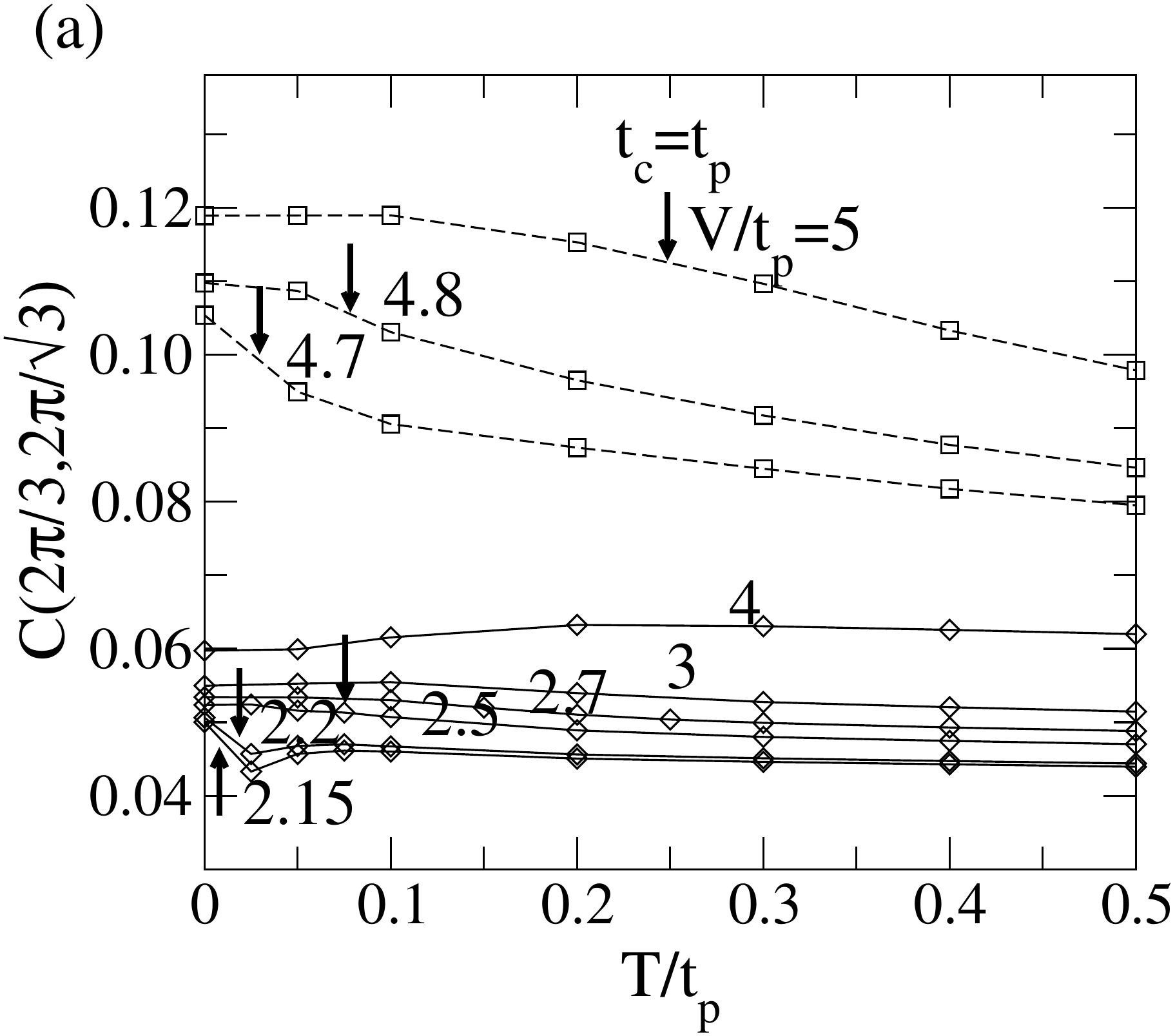}
     \includegraphics[clip,scale=0.23]{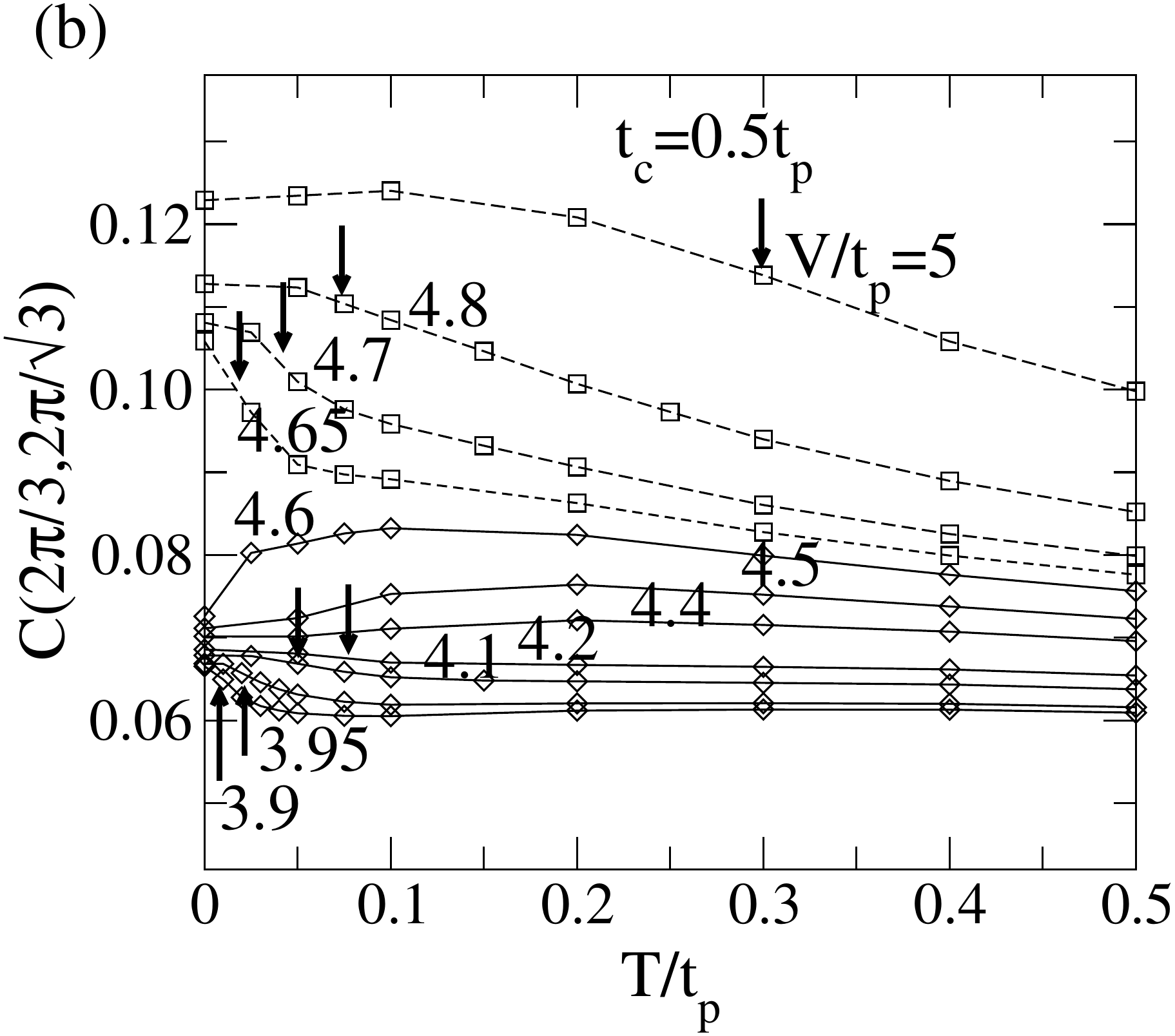}
     \includegraphics[clip,scale=0.23]{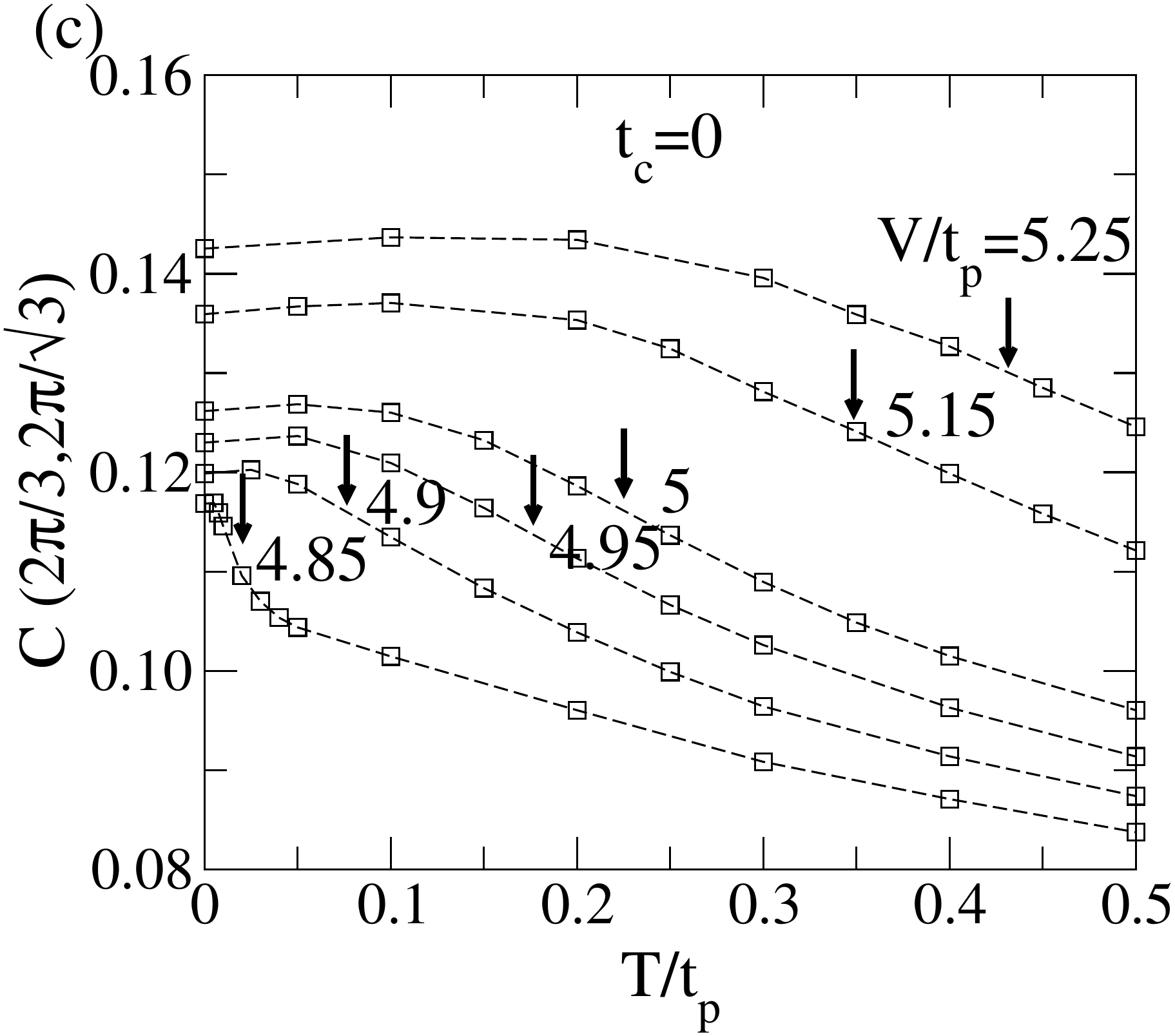}
     \includegraphics[clip,scale=0.23]{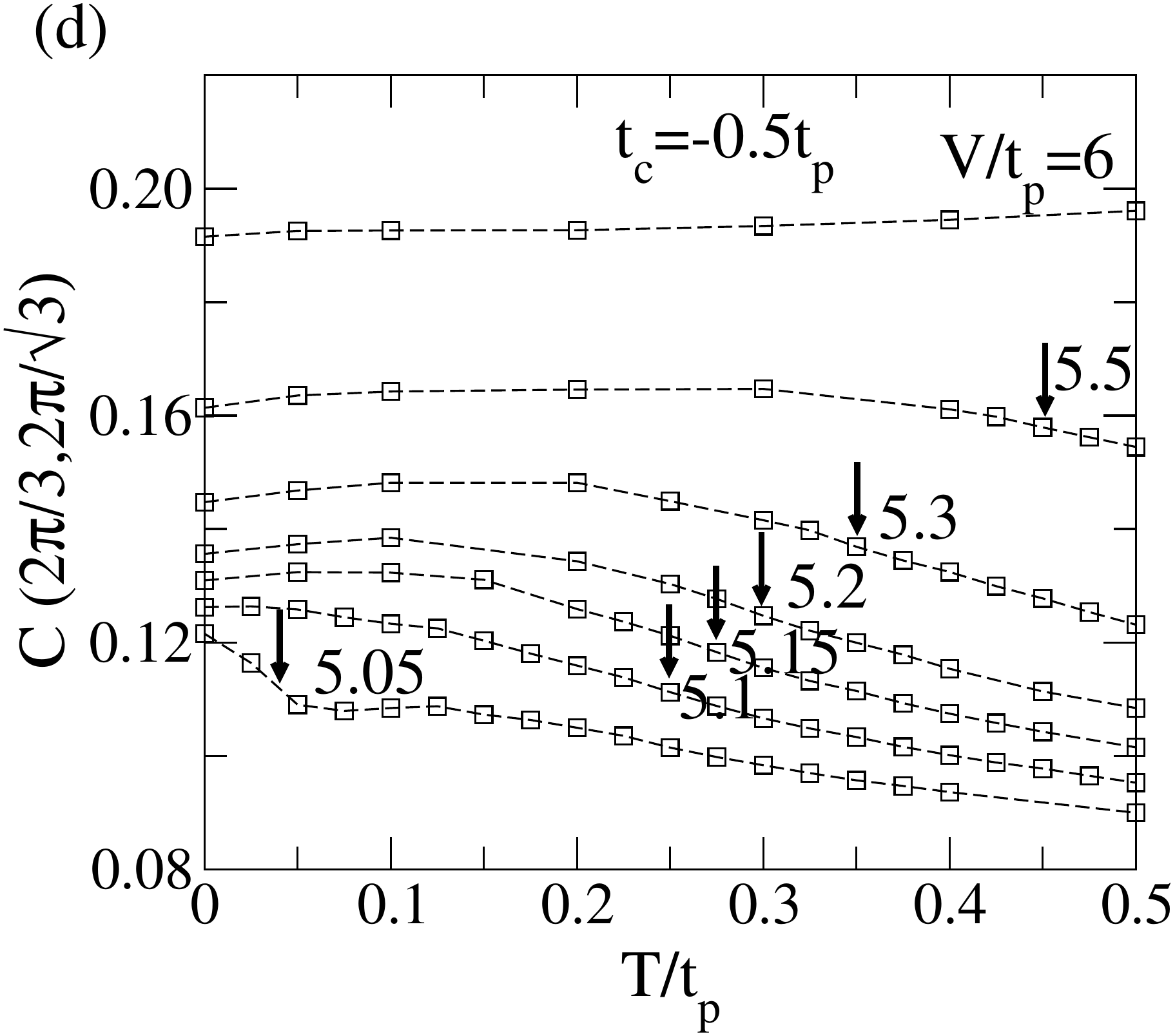}
     \includegraphics[clip,scale=0.23]{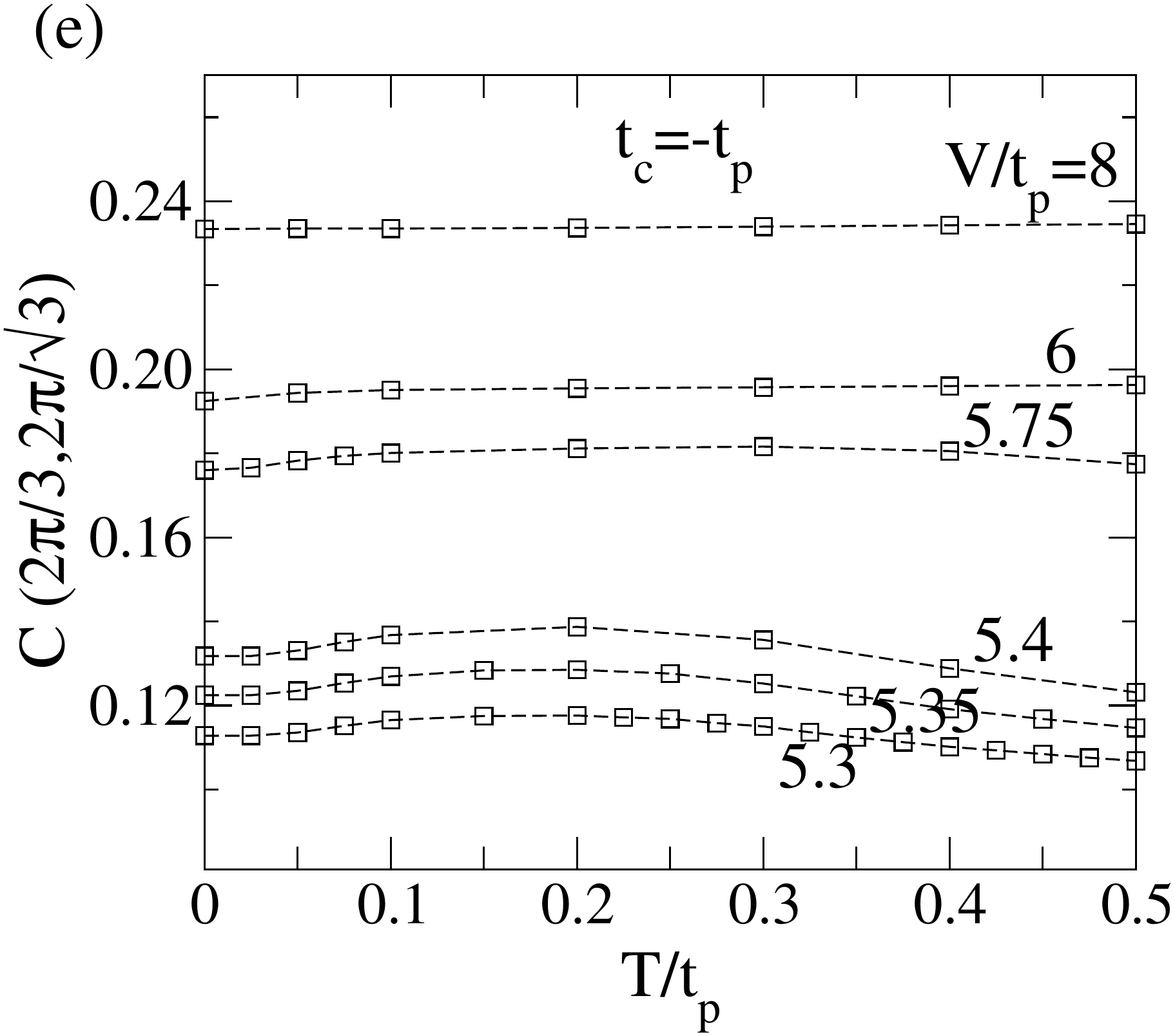}
     \caption{\label{CqU15V_T}
     Charge correlation function versus the temperature for different $V/t$ in the CO phase
     at $U=15t$ for $t_c/t_p=1,0.5, 0,-0.5,-1$
     Arrows correspond to inflection points of the curves, indicating
     the ordering temperature $T_{CO}$.}
\end{figure}

\begin{figure*}
 \center
      \includegraphics[clip,scale=0.65]{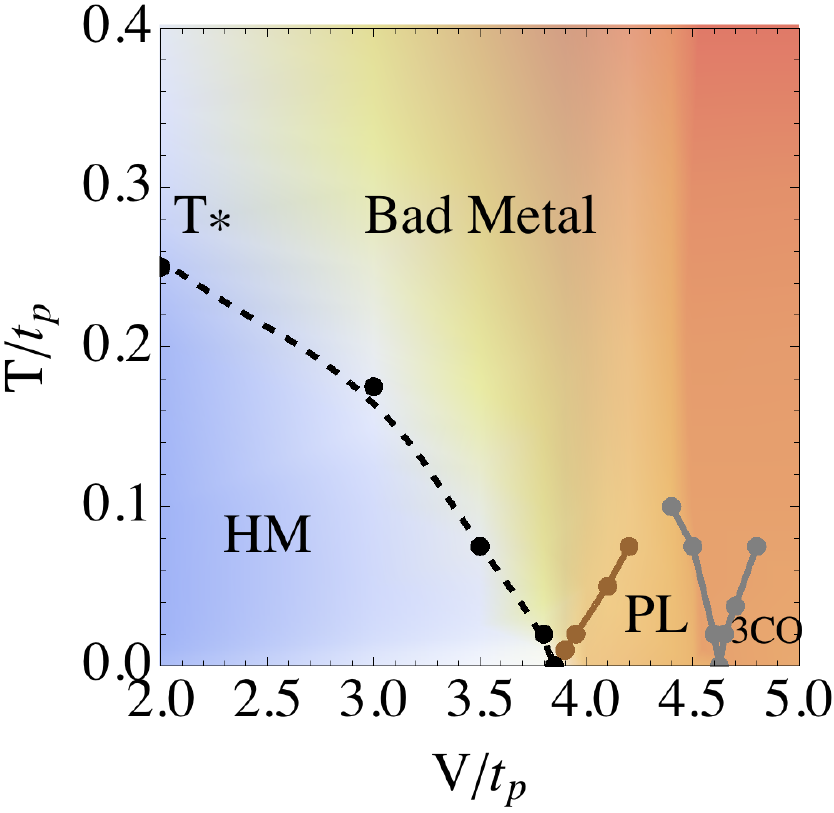}
     \includegraphics[clip,scale=0.65]{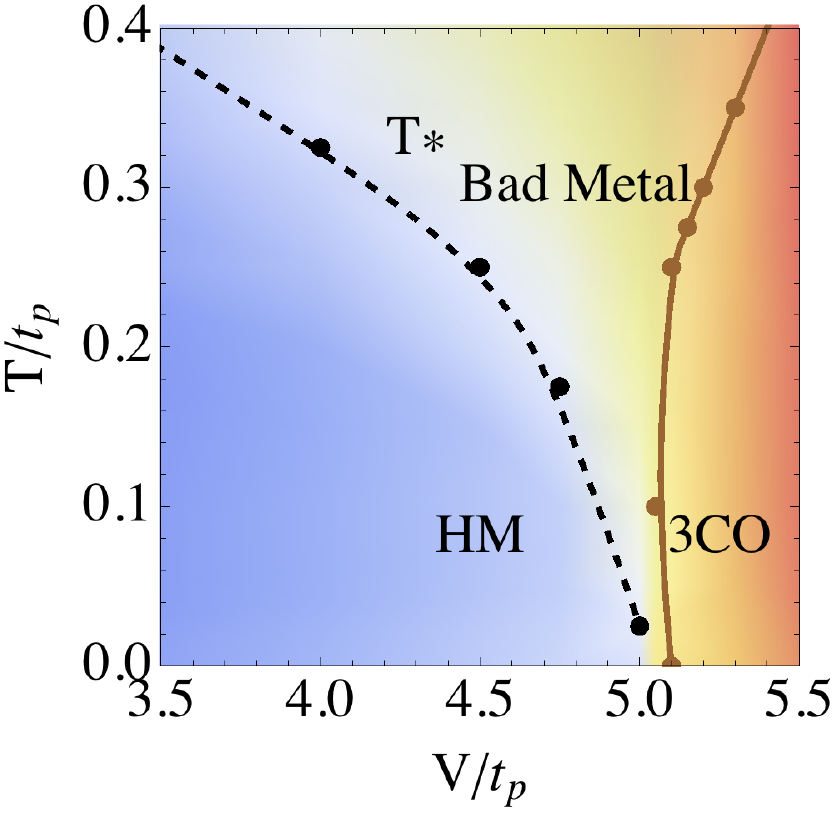}
     \includegraphics[clip,scale=0.65]{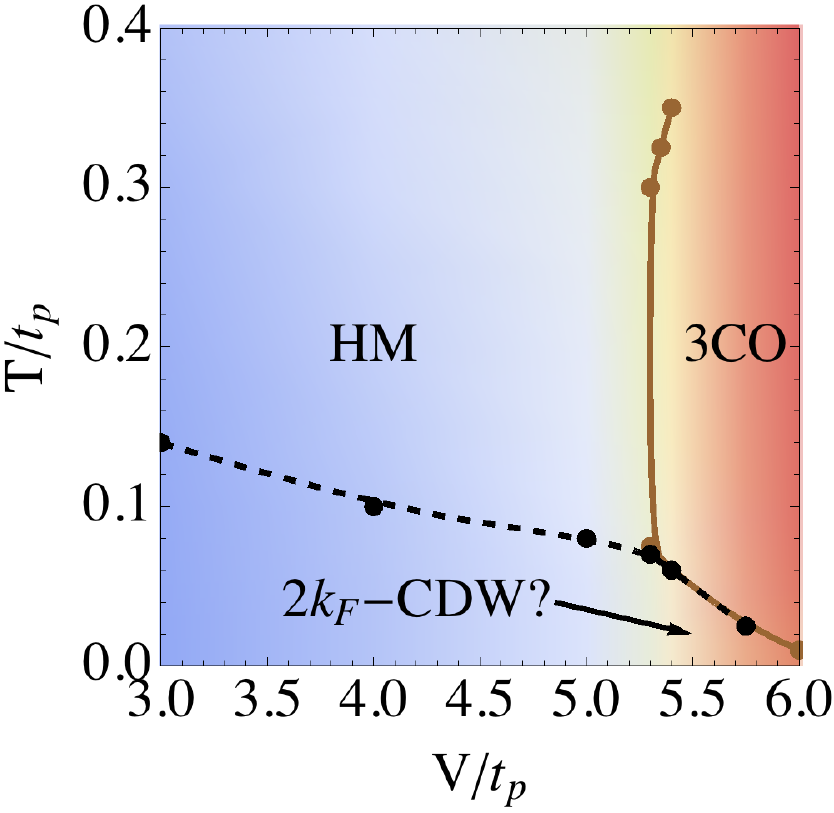}
     \caption{\label{fig:colorines} (Color online) Color plots summarizing the QCP
       behavior at $U=15t_p$ 
     for the representative cases 
     $t_c/t_p=0.5,-0.5$ and $-1$. The dashed curve is the renormalized Fermi
     temperature $T^*$ extracted from Fig. \ref{kineticU15V_T}. The full
     line is the charge ordering temperature $T_{CO}$ extracted from
     Fig. \ref{CqU15V_T}. The color gradients (blue and red
     respectively) are derived from
     the  kinetic energy and charge correlation data of
     Figs. \ref{kineticU15V_T} and \ref{CqU15V_T}.
} 
\end{figure*} 

The emerging QCP scenario can be further appreciated by studying the 
evolution of the  charge ordering transition $T_{CO}$ vs. temperature. This
can be obtained by tracking the steepest variation of the charge correlation
function, $C({\bf Q})$ (arrows in Fig.~\ref{CqU15V_T}a-d).
In Fig. \ref{fig:colorines}  we report $T_{CO}$
together with  the $T^*$ extracted from  figures \ref{kineticU15V_T},
showing a common behavior in  proximity
to the CO instability. As stated in the preceding paragraph, 
the case $t_c/t_p=-1$ exhibits a different behavior, with no visible
$T^*$ approaching the QCP. The charge correlations  (Fig. \ref{CqU15V_T}e) also 
exhibit a qualitatively different behavior in the 
perfectly nested case $t_c/t_p=-1$, with a mild
non-monotonic temperature dependence showing a maximum at intermediate temperatures which suggests 
a 'reentrant' ordering transition. For this case, there is no clear indication of 'bad' metallic
behavior in the kinetic energy and there is clear evidence of 'reentrant' behavior
in the 3CO transition. This 'reentrant' behavior disappears at around the critical value $V_c=6t_p$ as
can be noted in Fig. \ref{CqU15V_T}d.  It can be noted that a slightly reentrant behavior can be extracted
from  the data at $t_c/t_p=-0.5$.  The presence of a reentrant behavior at negative values of $t_c/t_p$
is confirmed by our RPA analysis, 
and is strongly reminiscent of what is commonly observed in
the EHM on the square lattice (cf. Fig. 1 in Ref. \cite{Merino06}),
which also in that case is ascribed to the competition of the CO phase
with a Fermi surface nesting instability.  We have added to the phase diagram for the case $t_c=-t_p$
the $2k_F$-CDW instability and may also be present in other cases. However, the limited
wavevector resolution of our small cluster calculation does not permit an accurate determination 
of the stability of the $2k_F$-CDW phase.  In fact, for $t_c=-0.5t_p$, we may also expect that 
 CDW instabilities occur in the proximity of the 3CO instability.

\subsubsection{Specific heat}

\begin{figure*}
 \center
  \includegraphics[clip,scale=0.3]{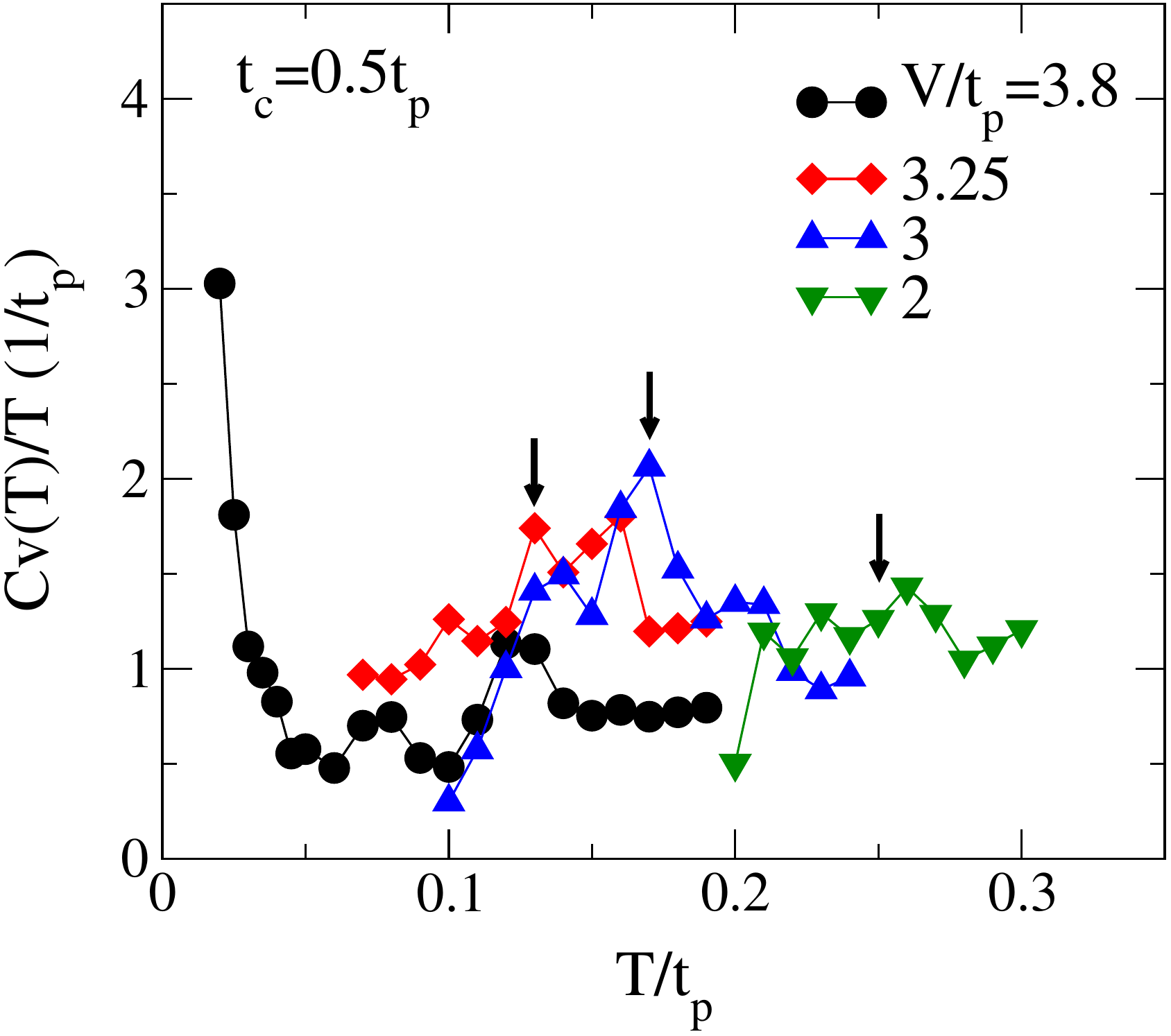}
  \includegraphics[clip,scale=0.3]{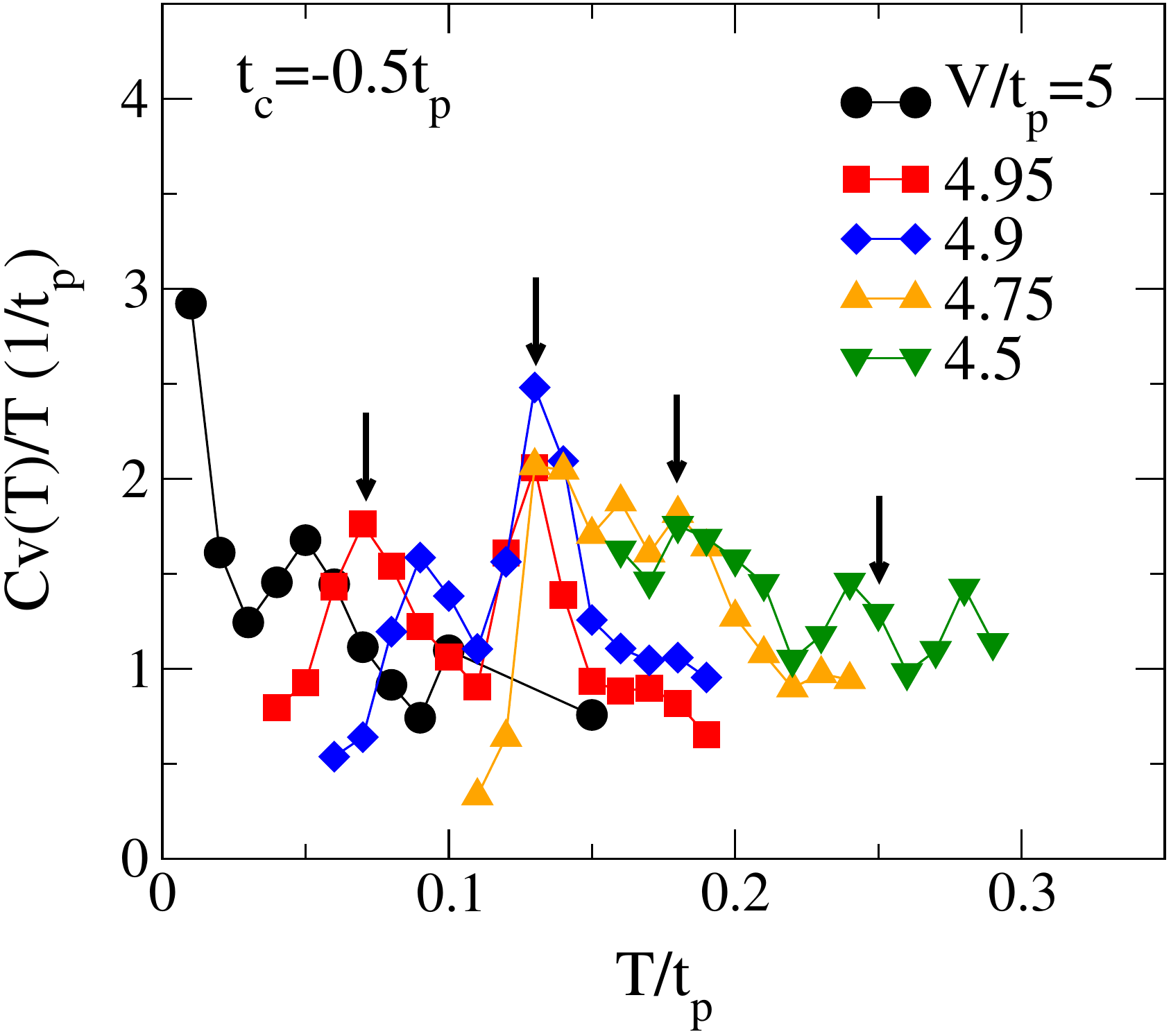}
    \includegraphics[clip,scale=0.3]{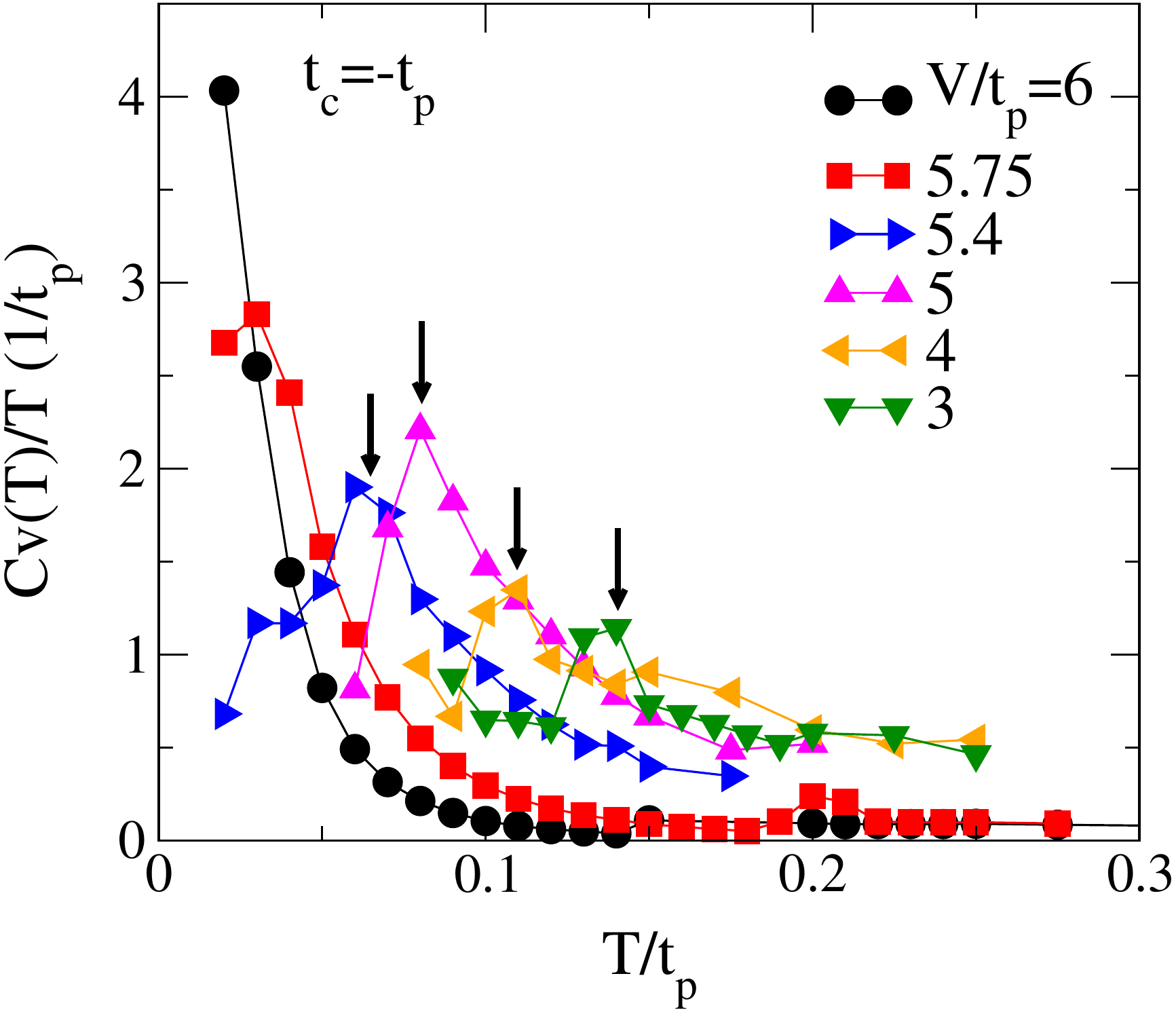}
  \caption{\label{fig:gamma}  (Color online) Temperature dependence of effective mass enhancement for different $t_c/t_p$ ratios for $U=15t_p$.
 A different behavior in the $T$-dependence is found for the case $t_c=-t_p$ for which $T^*$ (indicated by the vertical arrows)
 which indicates a departure from the quantum critical behavior found
 in the other non-nested Fermi surface situations.}
\end{figure*}

Further insight on the anomalous properties of the homogeneous phase
can be gained by exploring thermodynamic properties such as the temperature dependence of
the specific heat coefficient $C_V/T$ on approaching the QCP. In a Fermi
liquid at low $T$, this quantity measures the effective mass
enhancement of the quasiparticles.  In Fig. \ref{fig:gamma} we compare
$C_V/T$ for $t_c/t_p=0.5,-0.5, -1$.  In this way we compare
the behavior of the specific heat of a system across the 3CO ($t_c=-0.5t_p$) with a system across the PL transition ($t_c/t_p=0.5$). 
For completeness we also analyze the $t_c/t_p=-1$ case in which perfect nesting exists which can be compared to the
other two cases.  For both $t_c/t_p= \pm 0.5$  we find that a peak in $C_V/T$ develops at $T^*$
with $T^* \rightarrow 0$ on approaching the QCP as  $V \rightarrow V_c$.  This indicates that 
both the drop of $T^*$ and the effective mass enhancement occurring in proximity to the
 QCP are consistent with the phase diagram of Fig. 9.  Both effects are key signatures of the presence of 
a QCP together with the 'bad' metallic behavior arising around it. 

The case $t_c=-t_p$ deserves special attention. For Coulomb repulsion energies up to about $V=5.4t_p$, 
there is a moderate increase of the effective mass enhancements and a moderate shift of the peak to lower temperatures 
in contrast to previous cases. However, increasing $V$ further leads to different behavior with 
a rapid increase of the effective mass enhancement and a shift of the peak to zero 
which indicates the proximity to a 3CO transition at $V_c=6t_p$. The shift of the peak in the specific heat with $V$ is plotted in the phase diagram 
of Fig. \ref{fig:colorines} together with the transition line to the 3CO. The two lines merge and a clear 'reentrant'
behavior of the 3CO transition is observed which is ascribed to the presence of the competing Fermi surface nesting
instability.  The origin of the specific heat peak is unclear since it does not separate the HM from the 'bad' metal
and from Fig. \ref{fig:colorines} it is clear that the peak position has a different 
qualitative behavior for  $t_c/t_p=-1$  than for other $t_c/t_p$ ratios. In fact,  there is no 'bad' metallic behavior 
in the HM phase as discussed previously. However, the fact that the effective mass is enhanced may indicate
a transition to the $2k_f$-CDW.

\subsection{Connection with QCP physics}

 It is worth analyzing our numerical results from the perspective of the standard Moriya-Hertz-Millis (MHM)\cite{Hertz76,Millis93}
theory of quantum critical points (QCP). In principle, this theory could be appropriate to the transition from the HM to the 3CO phases
considered here since both are metallic.  However, one should keep in mind 
the limitations of the MHM theory. First, it is based on a weak coupling perturbative expansion around the
QCP.  Second, the MHM theory is not applicable for ordering 
transitions which are driven by ``2k$_f$'' 
Fermi surface instabilities. Hence, the MHM
theory is applicable only to systems in which there is no nesting at all and for which the ordering wavevector ${\bf Q}$ is not an extremal vector of the Fermi surface, {\it i. e.}  
loosely speaking $Q \ne 2k_f$.  Our model satisfies the latter
condition whenever $t_c/t_p \neq \pm 1$ 
for the 3CO wavevector: ${\bf
  Q}=(2\pi/3, 2\pi/\sqrt{3})$ 
(see Fermi surfaces and discussion in Appendix \ref{RPA}).  In these
cases, the two-dimensional version of MHM, $d=2$, with a dynamical
scaling dimension, $z=2$  
 (corresponding to nearly antiferromagnetic metals with effective dimension at  the QCP: $D=d+z=4$) is relevant. 
This situation corresponds to a marginal case which contains dangerously irrelevant operators in the renormalization group sense 
which can destroy the hyperscaling at the QCP.  In this marginal case, $d=z=2$,  the Fermi liquid phase is bounded in the phase diagram 
 by the condition: $T< r$, where $r=|V-V_c|$ quantifies the proximity to the QCP from the metallic side of 
 the transition and $V_c$ is the critical value at which CO occurs. 
 The quantum critical region is bounded by the condition $T>r$ obtained from the scaling behavior of the renormalization 
 group (RG) equations.  A similar 
linear dependence with $r$ is found in the boundary of the ordered phase which displays a critical 
non-Gaussian behavior around it  due to the Coulomb interaction.  
As summarized in Fig. \ref{fig:colorines}, for $t_c
/t_p\neq \pm 1$ 
the  suppression of $T^*$ at the QCP 
and the 'bad' metallic behavior obtained at finite-$T$ from
our numerical calculations 
are qualitatively consistent with the MHM predictions for the marginal
$d=z=2$ case.

On the other hand, the extremal values: $t_c/t_p=\pm 1$ deserve special consideration. In the particular case: $t_c=-t_p$, 
the system has perfect nesting at the ordering wavevector: ${\bf Q}_F=(\pi,\pi/\sqrt{3})$,  (see Fig. \ref{fig:FS} in Appendix \ref{RPA}) 
which describes diagonal stripe order in real space. A
RPA analysis  on the model shows that
 a CDW instability at  ${\bf Q}_F$ exists at
small but finite $U$ and $V$, which competes  
with the 3CO with ${\bf Q}=(2\pi/3, 2\pi/\sqrt{3})$. Such coexistence/competition
between CDW and CO instabilities has also been found in RPA studies of the extended Hubbard model on the 
square lattice \cite{Merino06}. 
The situation in which two instabilities coexist ---a nesting driven CDW with ordering vector ${\bf Q}_F$ and a Coulomb driven
 instability with ordering vector ${\bf Q}$--- has not been
 addressed in general at the level of the MHM approach.
 Our analysis shows that at moderate 
 values of $U\gtrsim 7t_p$  nesting instabilities are
 washed away and the charge ordering transition is  Coulomb driven.  
However, the competition washes out the $T^*$
phenomenon and we find no clear evidence of 'bad' metallic
behavior around the QCP in the  case $t_c=-t_p$.
This coincides with the breakdown of  
the MHM approach when nesting is present in the lattice.

We may speculate, based on our numerical analysis, that in the perfectly nested situations quantum criticality is destroyed and
a first order transition occurs. Indeed, a somewhat related renormalization group (RG) approach
\cite{Altshuler95} to $2k_F$-density wave
 quantum phase transitions in which curved Fermi surfaces
with parallel tangents at two points of the Fermi surface connected by $2k_F$ are considered has found that critical fluctuations strongly influence the 
fermions on the Fermi surface and that the feedback effect of these fluctuations can destroy the second order quantum critical point 
turning it into a first order transition. Only in the special case in which ${\bf Q}={\bf G}/2$, with ${\bf G}$ being a
reciprocal lattice vector, a second order quantum phase transition is
recovered.

In actual quarter-filled organic materials, $\theta$-ET$_2$X, for which hopping ratios $t_c/t_p \ne -1$ 
the MHM theory may be relevant. Many of the predictions for thermodynamic 
and transport properties in the  quantum critical regime above the zero temperature QCP could be 
then experimentally checked. One important prediction of the MHM theory  for  
$d=z=2$ is the anomalous temperature dependence of the specific heat : 
  \begin{eqnarray}
 C &\sim&  T {\partial S  \over \partial T } |_V \sim T\ln{1/r}, \;\; T\ll r\\ 
   & & T \ln{1/T}, \;\; T\gg  r,
 \label{Cv}
 \end{eqnarray}
 where $r$ describes the proximity to the QCP.  
On the other hand, in clean nearly charge ordered two-dimensional metals the resistivity around the 'hot' spots in the 
quantum critical regime\cite{Lohneysen,Rosch1997} reads:
\begin{eqnarray}
\rho &\propto& T^2,\;\; T\ll r \\
            & & T,\;\;  T\gg r.
\label{rho}            
\end{eqnarray}
However, the  resistivity is shortcircuited by the contribution of electrons at the 'cold' spots 
since the scattering is small around these parts \cite{Hlubina1995}. Hence, the non-Fermi liquid behavior 
at the 'hot' spots is masked by the 'cold' sections eventually restoring 
Fermi liquid behavior:  $\rho \sim T^2$.
 Therefore, within the MHM approach 
 and in the the quantum critical region, the specific heat coefficient displays divergent 
 behavior as $T \rightarrow 0$ following Eq. (\ref{Cv}). The resistivity could show non-Fermi liquid behavior
under small disorder which has been found to strongly influence antiferromagnetic QCP's \cite{Rosch1999}. 
Averaging the scattering rate over the Fermi surface reduces the effectiveness of the Hlubina-Rice
mechanism and the scattering from the 'hot' regions becomes effective leading again to 
non-Fermi liquid behavior \cite{Rosch1999}.

The behavior of the specific heat coefficient that we have found around the critical point (see Fig. \ref{fig:gamma}) does show 
an enhancement on approaching the QCP in consistent agreement with Eq. (\ref{Cv}). However, we cannot accurately
determine  the logarithmic dependence from our numerical data due to the small cluster sizes reached.

\section{Conclusions and outlook}
\label{sec:conclusions}

We have analyzed in detail the effect of geometrical frustration on
charge ordering  
transitions realized in the 
extended Hubbard model on 
the anisotropic triangular lattice, which appropriately describes the
family of quarter-filled layered organic crystals: 
$\theta$-(ET)$_2$X.  The model contains both  
onsite, $U$, and inter-site Coulomb repulsion terms  $V_p$  and
$V_c$, that are taken to be isotropic, 
$V_c=V_p=V$.
The degree of geometrical frustration in the 
electron motion
is tuned through the $t_c/t_p$ ratio, which is an important parameter
controlling  the experimental phase diagram.

The zero temperature phase diagram of this model contains a homogeneous metal (HM),  a pinball liquid (PL) and
a three-fold charge ordered (3CO) phase.  While the 3CO phase occurs
at sufficiently  strong inter-site interactions $V$
for any fixed $U$, the PL only occurs above a certain threshold $U$
value, as its existence is inherently tied up 
to the strong coupling regime.  
On the other hand, the PL is found to be stabilized by increasing the geometrical frustration of the lattice. 
Our results do show that in the range of values of $U/t_p$ and $V/t_p$ appropriate 
to the $\theta$-(ET)$_2$X  materials, increasing the geometrical frustration of the lattice can effectively 
tune the system from a homogeneous metal (HM) with strong charge order correlations to a ``pinball" liquid (PL) phase.

The phase transitions between charge ordered  and disordered metallic phases 
can display quantum critical phenomena in close analogy with the heavy
fermion systems  
\cite{Coleman,Gegenwart,Lohneysen,Miyake} with the critical charge rather than the spin fluctuations driving the
CO transition.  Such type of fluctuations may be at the origin of both the anomalous properties in the metallic state and Cooper-pair formation. 
 Indeed, non-Fermi liquid behavior as well as non-BCS
superconductivity have both been predicted and observed in
quarter-filled organic 
materials of the $\alpha$, $\beta''$ and $\theta$-(ET)$_2$X type \cite{Ishiguro,Takenaka1,Yasuzuka,Weng,Merino01}.
Such heavy fermion behavior arising from molecular $\pi$ electrons
instead of the $d$ or $f$ electrons, 
as occurs in the rare earths, may indeed find a natural explanation based on
the properties of matter expected near a QCP.

In order to establish whether quantum critical behavior occurs or not
in the quarter-filled layered materials close to CO  several issues could be experimentally and theoretically addressed: 
(i) Is there evidence for the divergence of the specific heat
coefficient and the quasiparticle effective mass, 
$m^*/m \rightarrow \infty $  and for the collapse 
 of the Fermi temperature, $T^* \rightarrow 0$,  near the QCP? 
Measurements of the quadratic coefficient of the 
 resistivity approaching  the QCP from the Fermi liquid side of the critical point can be useful to test the effective mass enhancement.  
 Such type of experiments have been systematically performed in $\kappa$-(DHDA-TTP)$_2$SbF$_6$ and
(MeDH-TTP)$_2$AsF$_6$, by tuning the system across the CO transition
via applied pressure  \cite{Yasuzuka,Weng}, yielding phase diagrams similar to those of Fig.
\ref{fig:colorines}.
(ii) Is there non-Fermi liquid behavior of transport and thermodynamic properties  
 in the quantum critical regime above the QCP? What is the temperature dependence 
 of  the resistivity in these systems? Are there clear deviations from Fermi liquid behavior of the form in Eqs. (\ref{Cv}) and (\ref{rho})? (iii) If quantum criticality and scaling are 
 observed in transport and thermodynamic quantities, how much of this behavior is consistent with the MHM predictions?
 Could there be a new universality class around the QCP in
 quasi-two-dimensional organic materials, related to the emergence of
 the pinball phase? (iv) Measurements
 of the Hall coefficient can be useful to disentangle whether the QCP is of the MHM type or different.  In standard 
 MHM theories, the Fermi surface would fold due to Bragg reflection
 off the density wave with no discontinuity in the Hall constant 
 when the system is tuned across the QCP\cite{Coleman,Gegenwart}.
 However, as in heavy fermions a local type of QCP could arise in
 which the system jumps discontinously 
 from a large Fermi surface to a small Fermi surface through the QCP leading to a discontinuous jump of the Hall coefficient.  Since the transition
from the HM to the PL  involves localization of the 'pin' electrons,  a transition from the large Fermi surface of the HM
involving all carriers to a small Fermi surface involving only 'ball'
itinerant electrons   could indeed occur. 
Understanding how this transition takes place and the type of QCP observed could be 
resolved by Hall constant measurements in analogy to the heavy
fermion systems. (v) Here  we have mainly discussed transitions 
between disordered and ordered {\it metallic} phases for which MHM
theory is meant for. An important issue to  address is how  quantum criticality is modified in transitions
from HM to CO  {\it insulating} phases? This issue can be addressed
within the EHM studied in the present work, by allowing for
anisotropic Coulomb interactions $V_p\neq V_c$. 

As observed in Ref. \cite{Mori06} the superconductivity in $\theta$-(ET)$_2$X
compounds, as in other polytypes, frequently appears near to an insulator. In such cases, the cause of
superconductivity (SC) may not be the simple weak coupling BCS mechanism by the electron-phonon interaction, but rather due to electronic
correlations. Several theoretical works in the weak coupling limit have been
performed in order to examine the 
possible mechanism for the onset of SC in proximity to the CO phase
\cite{Scalapino,Merino01,Tanaka}. Unconventional SC of the $f$-wave type 
has been encountered on the anisotropic triangular lattice with the model parameters: $t_c=0$ and $V_c=V_p \approx t_p $   \cite{WatanabeVMC06}
with $U=10t_p$ and mediated by the charge fluctuations. 
This $f$-wave pairing symmetry is the analogous to the $d_{xy}$-wave pairing found in proximity to the checkerboard CO on
the square lattice \cite{Merino01}. It would be interesting to search for unconventional SC around the QCP found
for other $t_c/t_p \neq 0 $ ratios and $U$ values both in the 3CO and  PL  type QCP.  Based on the results of the present
work,  it can be conjectured that the anomalous properties and unconventional superconductivity  observed in
X$=$I$_3$ compounds maybe related to the proximity to the strong coupling PL phase since the onsite
Coulomb repulsion energy is significant: $U=(15-20)t_p$.  This could be tested by applying uniaxial pressure on 
the $\theta$-(ET)$_2$I$_3$ crystals.

Other materials such as the rare-earth nickelates, AgNiO$_2$, do show three-fold CO transitions similar to the
one discussed here \cite{Coldea11} although the origin of the CO transition may be non-Coulomb
in origin since AgNiO$_2$ has a complex multiorbital structure \cite{Mazin07} and other effects such
as Hunds coupling and crystal fields can play a relevant role. Thus, the organic materials of the $\theta$-(ET)$_2$X  
type appear to be ideal candidates to single out the effects on electronic properties of metals close 
to a charge-driven quantum phase transition mediated solely by the offsite Coulomb repulsion 
between electrons at different sites.

\appendix
\section{RPA results at weak $U$}
\label{RPA}

\begin{figure}
 \center
    \includegraphics[scale=0.45]{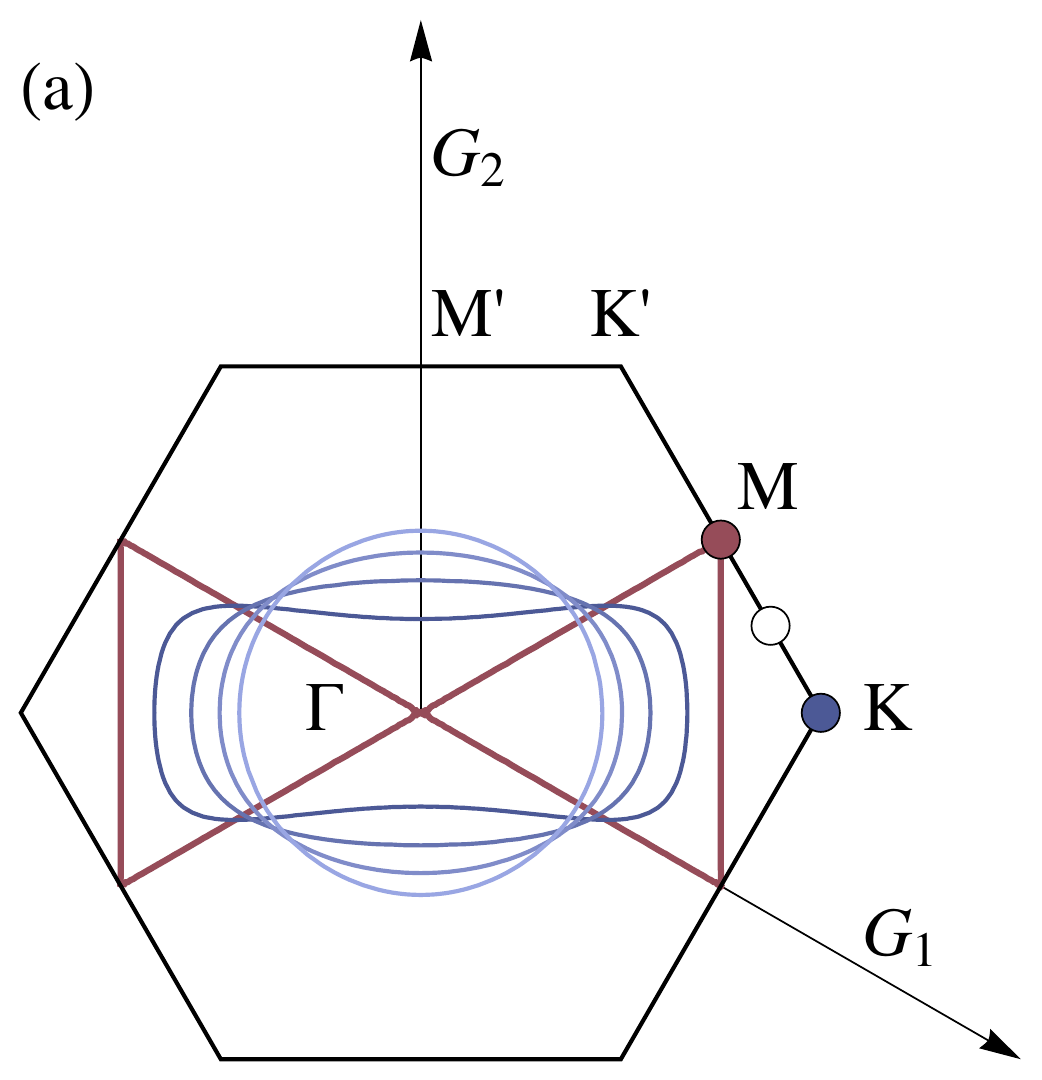}
   \includegraphics[clip,scale=0.4]{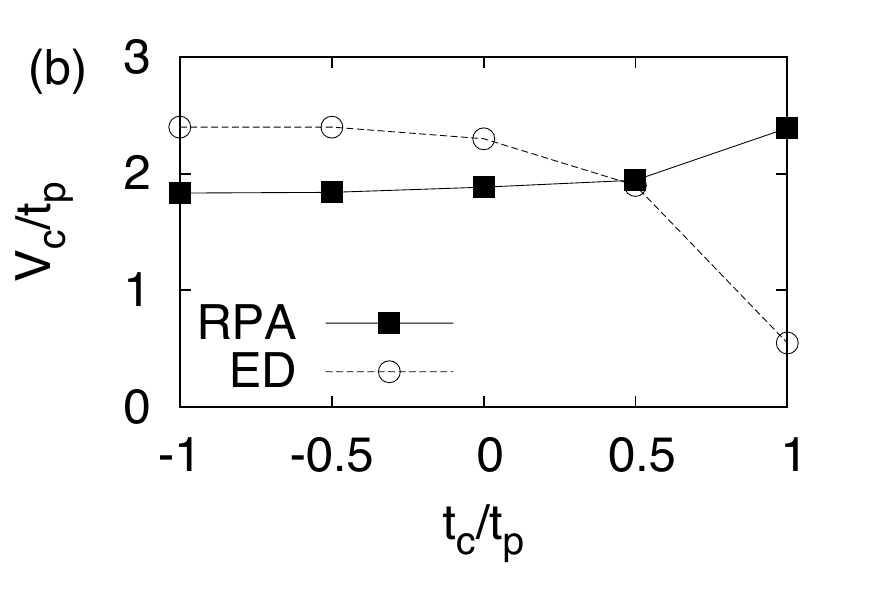}
   \includegraphics[scale=0.7]{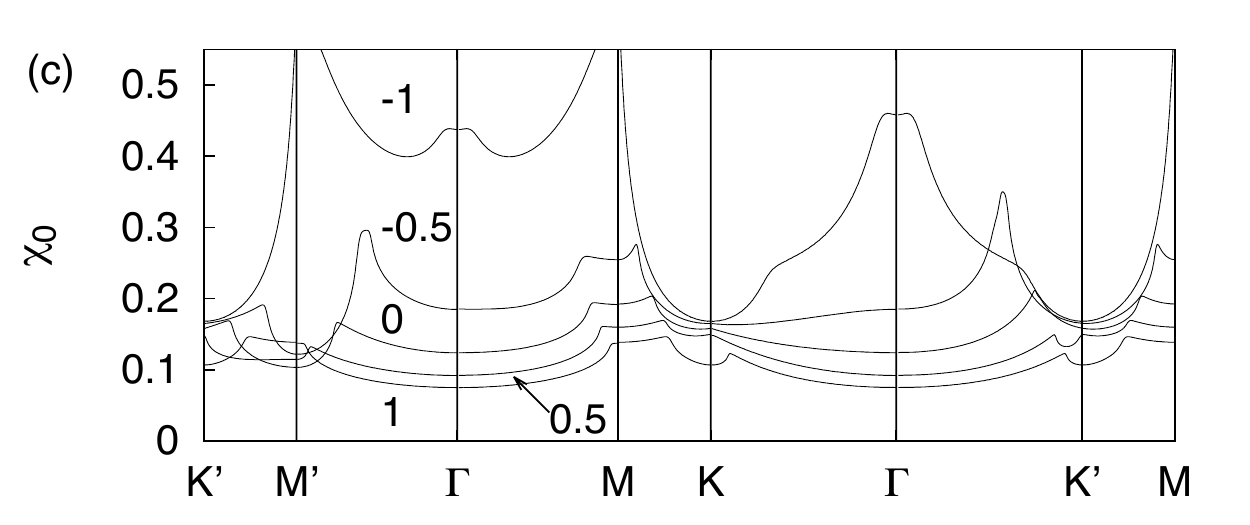}
   \caption{\label{fig:FS} (Color online) (a) Brillouin zone of the triangular
    lattice (black), reciprocal lattice vectors
    $G_1=(2\pi,-2\pi/\sqrt{3})$ and $G_2=(0,4\pi/\sqrt{3})$ (arrows),
    and   Fermi surface: from light blue to dark blue,
    $t_c/t_p=1,0.5,0,-0.5$; red, $t_c/t_p=-1$ 
    (same color code as in Fig. \ref{phaseVU_T0}). The blue and red
    dots are respectively the threefold wavevector 
    ${\bf Q}={\bf K}=(4\pi/3,0)$ and the nesting wavevector 
    ${\bf Q}_{F}={\bf M}=(\pi,\pi/\sqrt{3})$ connecting flat segments of the
    Fermi surface at $t_c/t_p=-1$ (red), while the white dot is the wavevector 
    associated with the predominant instability for $t_c/t_p=1$;  
(b) the critical value
   of the intersite Coulomb interaction $V_{3CO}$ for $U=5t_p$,
   comparing the RPA and ED results.
(c) Free-electron susceptibility $\chi_0$ along symmetry lines of the Brillouin zone, for different values of $t_c/t_p$.
}
\end{figure}

In the random phase approximation (RPA), the instability of the
homogeneous metal is signalled by a divergence of the charge
susceptibility\cite{Merino06} 
\begin{equation}
  \label{eq:chi}
  \chi({\bf q})=\frac{\chi_0({\bf q})}{
 1+[U/2+V({\bf q})]\chi_0({\bf q})}
\end{equation}
at a given wavevector. Here $\chi_0({\bf q})$ is the non-interacting
susceptibility of the lattice,  $V({\bf q})$ is the interaction
potential in Fourier space and $U$ is the onsite repulsion. 
For isotropic n.n interactions on the
triangular lattice we have 
\begin{eqnarray}
  \label{eq:pot}
  V({\bf q})&=&2V\left\lbrace \cos (q_x) +\right. \\
& + &  \cos[(q_x+\sqrt{3}q_y)/2]+\cos[(q_x-\sqrt{3}q_y)/2]
\rbrace. \nonumber
\end{eqnarray}
An instability occurs when the denominator in Eq. (\ref{eq:chi})
vanishes, which requires $-V({\bf q})=\chi^{-1}_0({\bf q})+U/2$. In
principle the above equation can describe both {\it charge ordering}, driven
by the Coulomb iteraction $-V({\bf q})$ that is maximum at the six equivalent
threefold wavevectors  ${\bf Q}=(\pm 2\pi/3,\pm 2\pi/\sqrt{3})$, 
$(\pm 4\pi/3,0)$  (blue dot in
Fig. \ref{fig:FS}a), and a charge density wave (CDW) induced by a
large $\chi_0$. The evolution of the free-electron susceptibility with
frustration is illustrated in Fig. \ref{fig:FS}b.

A CDW instability occurs for $t_c/t_p=-1$ 
due to a perfect nesting between parallel
segments of the Fermi surface (red lines in
Fig. \ref{fig:FS}a), at wavevector  ${\bf Q}_{F}=(\pi,\pi/\sqrt{3})$
(red dot, corresponding to the {\bf M}-point), 
but is washed out at $U/t_p\gtrsim 7$, where the threefold
order is always favored.
An  instability also appears to compete with the 3CO 
 in the case $t_c/t_p=1$ for 
$U\lesssim 2 t_p$, at a wavevector ${\bf Q}_1=(2.62,2.72)$ (white dot). 
Such vector lies at the intersect between a circle of radius 
$|{\bf Q}_1|=2k_F$ and the boundary of the Brillouin zone. 
It represents a compromise between a Fermi surface instability and a 
genuine charge ordering, as it benefits from both a large $\chi_0$ and
a large $-V({\bf q})$. For values of the frustration ratio
$|t_c/t_p|\lesssim 0.9$, the RPA predicts that the 3CO transition is
dominant for all $U>0$. 

The critical coupling $V_{3CO}$ at the threefold instability is shown as a
function of the geometrical frustration in
Fig. \ref{fig:FS}b. The RPA predicts an increase of $V_{3CO}$ with
$t_c/t_p$ which originates from a decrease of the density of states at
the Fermi level \cite{Nishimoto09}.
We see that even at a relatively low value of $U=5t_p$, 
the  RPA result does not agree with the exact
diagonalization data, showing an opposite trend for positive $t_c/t_p$
ratios. In the ED, the $t_c$ dependence 
is governed by the stabilization of the 
pinball liquid phase, that is not captured by the weak coupling RPA argument.

\section{Mean-field potential energy  in the CO state}
\label{Hartree}
The Hartree expression for the potential
energy per site  in a charge ordered state with three-fold symmetry  reads:
\begin{eqnarray}
E_{H}&=&{U \over 3} (n_{A\uparrow}n_{A\downarrow}+n_{B\uparrow}n_{B\downarrow} + n_{C\uparrow}n_{C\downarrow})
\nonumber \\
&+&V(n_An_B+n_An_C+n_Bn_C).
\end{eqnarray} 
We take $n_{A\uparrow}=n_{A\downarrow}=
n_{B\uparrow}=n_{B\downarrow}=1$, $n_{C\uparrow}=1/2,n_{C\downarrow}=0$ 
corresponding to the Hartree solution for the 3CO state and
$n_{A\uparrow}=n_{B\uparrow}=3/4$, 
$n_{A\downarrow}= n_{B\downarrow}=1$, $n_{C\uparrow}=1,n_{C\downarrow}=0$
for the pinball liquid phase (see Fig.\ref{fig:dens}). 
Inserting these values in the preceding expression  we find:  
\begin{eqnarray}
E_H^{(M)}&=&(9/16)U+ (27/4)V, \;\;\; \text{HM}
\label{EM}  \\
E_H^{(PL)}&=&(1/2) U+ (13/2+1/16) V, \;\;\; \text{PL}
\label{EPL} \\
E_H^{(3CO)}&=&(2/3) U+  6V, \;\;\; \text{3CO}
\label{E3CO}
\end{eqnarray}
The potential energy calculated by ED 
closely follows the $V$ dependence predicted by the above mean-field 
equations (not shown). 
The transition from the pinball state to the three-fold CO state is well
captured by the mean-field analysis: the ED data closely follow the
value $V_c=U/3$ obtained by equating $E_H^{(PL)}=E_H^{(3CO)}$.

\section*{Acknowledgements.} 
We acknowledge P. Horsch, A. Greco, A. Liebsch and M. Tamura for fruitful discussions. 
J.M. and L.C. acknowledge financial support from
MICINN (CTQ-2008-06720-C02-02 and Consolider CSD2007-00010).

\end{document}